\documentclass{aa}  

\usepackage{graphicx}
\usepackage{txfonts}
\DeclareRobustCommand{\VAN}[3]{#2}
\let\VANthebibliography\thebibliography
\def\thebibliography{\DeclareRobustCommand{\VAN}[3]{##3}\VANthebibliography}

\usepackage{amsmath}	% Advanced maths commands
\usepackage{amssymb}	% Extra maths symbols
\usepackage{adjustbox}
\usepackage{array}
\usepackage{lscape}
\usepackage{longtable}
\usepackage{float}
\usepackage{caption}
\usepackage{etoolbox}
\usepackage{appendix}
\usepackage{chngpage}
\usepackage{calc}
\usepackage{tablefootnote}
\usepackage{caption}

\begin{document}

%%%%%%%%%%%%%%%%%%%%%%%%%%%%%%%%%%%%%%%%%%%%%%%%%%

%%%%% AUTHORS - PLACE YOUR OWN COMMANDS HERE %%%%%

% Please keep new commands to a minimum, and use \newcommand not \def to avoid
% overwriting existing commands. Example:
%\newcommand{\pcm}{\,cm$^{-2}$}	% per cm-squared

%%%%%%%%%%%%%%%%%%%%%%%%%%%%%%%%%%%%%%%%%%%%%%%%%%

%%%%%%%%%%%%%%%%%%% TITLE PAGE %%%%%%%%%%%%%%%%%%%

% Title of the paper, and the short title which is used in the headers.
% Keep the title short and informative.
\title{The turbulent life of NGC~4696 as told by its globular cluster system\thanks{This paper includes data gathered with the $6.5$ meter \textit{Magellan} Telescope located at Las Campanas Observatory, Chile, and obtained with MegaCam (\citealt{Megacam}). The data products are produced by the OIR Telescope Data Center, supported by the Smithsonian Astrophysical Observatory. The full catalog of GC candidates (Table~\ref{tab:catalog}) is only available in electronic form at the CDS via anonymous ftp to cdsarc.u-strasbg.fr (130.79.128.5) or via http://cdsweb.u-strasbg.fr/cgi-bin/qcat?J/A+A/}}
\titlerunning{The turbulent life of NGC~4696}

% The list of authors, and the short list which is used in the headers.
% If you need two or more lines of authors, add an extra line using \newauthor
\author{
S. Federle\inst{1,2}\thanks{E-mail: sara.federle@eso.org}, M. G{\'o}mez\inst{1}, S. Mieske\inst{2}, W.~E. Harris\inst{3}, M. Hilker\inst{4}, I.~A. Yegorova\inst{2} \& G.~L.~H. Harris\inst{5}}
\authorrunning{S. Federle et al.}
% List of institutions
\institute{$^1$Facultad de Ciencias Exactas, Universidad Andres Bello, Chile \\
$^2$European Southern Observatory, Alonso de Cordova 3107, Vitacura, Santiago, Chile \\
$^3$Department of Physics and Astronomy, McMaster University, Hamilton, ON~L8S~4M1, Canada \\
$^4$European Southern Observatory, Karl-Schwarzschild-Strasse 2, 85748 Garching bei M{\"u}nchen, Germany \\
$^5$Department of Physics and Astronomy, Waterloo Institute for Astrophysics, University of Waterloo, Waterloo, ON N2L 3G1, Canada \\
}

% These dates will be filled out by the publisher
\date{Accepted XXX. Received YYY; in original form ZZZ}

% Enter the current year, for the copyright statements etc.

% Don't change these lines

\abstract
{Globular clusters (GCs) are remarkable witnesses of their host galaxy’s interaction and merger history.}{Our aim is to perform the photometric analysis of the globular cluster system (GCS) of the giant elliptical NGC~4696, which is the brightest member of Centaurus, a rich and dynamically young galaxy cluster.}{We obtained deep \textit{Magellan} $6.5$~m/MegaCam ($g'$, $r'$, $i'$) photometry, with which we identified a sample of $3818$ stellar clusters around NGC~4696 that were analyzed in the context of possible interactions and its assembly history.} {After carefully modeling and subtracting the galaxy light, we used selection criteria based on the shape, colors, and magnitudes to identify GC candidates. We find a number of features that indicate a disturbed GCS that points toward a complex evolution with other neighboring members of Centaurus. Formally, two subpopulations could be found at $(g'-i')_0 = 0.763\pm 0.004$ and $(g'-i')_0=1.012\pm 0.004$. Moreover, the color distribution does not show the presence of a significant blue tilt, but it presents a trend with the radius, where at small galactocentric distances a unimodal distribution is preferable to a bimodal one, suggesting the presence of an intermediate GC population. Besides the color distribution, the metallicity distribution also shows a bimodal trend, with peaks at $[\rm{Fe/H}]=-1.363\pm 0.010$ and $[\rm{Fe/H}]=-0.488\pm 0.012$. The radial density profiles show different slopes for the blue and red populations and the azimuthal distributions are well fitted by an asymmetrical sinusoidal function, with peaks projecting toward two nearby galaxies, NGC~4696B and NGC~4709, indicating past interactions among these three galaxies. Finally, we derived a GC specific frequency of $S_N=6.8\pm 0.9$, in good agreement with the values obtained for other giant ellipticals and with previously estimated $S_N$ of NGC~4696.}{All these results point toward a complex GCS, strongly influenced by the interaction history of NGC~4696 with the other galaxies of the Centaurus cluster. In a future work, the spectroscopic follow-up of the GC candidates analyzed in this study and broadening the photometric baseline will allow us to highlight the formation and evolution of the entire Centaurus cluster. }

\maketitle
%%%%%%%%%%%%%%%%%%%%%%%%%%%%%%%%%%%%%%%%%%%%%%%%%%

%%%%%%%%%%%%%%%%% BODY OF PAPER %%%%%%%%%%%%%%%%%%

\section{Introduction}

In the last decades, the study of globular clusters (GCs) has advanced significantly. They are defined as compact, gravitationally bound systems of stars, which have a typical mean mass of $\sim 2\times 10^5$~M$_{\odot}$ (\citealt{Beasley}). Even though we are missing some GCs in the region of the Galactic center (\citealt{Minniti21}) and disk (\citealt{Binney}), the study of the globular cluster system (GCS) of the Milky Way can be used as a template for the analysis of extragalactic GCs. In particular, the GCS of the Milky Way contains $~170$ GCs (\citealt{Ishchenko23}), with a large halo population (\citealt{Harris10}). They show a bimodal color (e.g., \citealt{Renaud17}) and  metallicity distribution, with the metal-rich GCs lying within $10$~kpc from the Galactic center, whereas the metal-poor GCs extend up to $145$~kpc in the Galactic halo (\citealt{Beasley}). 
In the Milky Way, we can resolve GCs into their single stars, whereas as we go outside the Local Group they appear as point-like sources. From the study of GCSs in other galaxies (e.g., \citealt{Racine}; \citealt{Puzia}), two scenarios seem to emerge: the presence of a single or of different populations of GCs highlighted from the uni-, bi-, or multimodality of their color distribution. In the case of NGC~1277, a single population of red GCs was observed, suggesting that the galaxy has undergone little or no accretion after its initial collapse (\citealt{Beasley18}). On the other hand, the GCS of NGC~4365 for which a trimodal color distribution was identified with blue, "green", and red GC populations (\citealt{Puzia}). Other examples are the GCS of NGC~4382, for which, besides two old GC populations, a young one with an age of $2.2\pm 0.9$~Gyr was observed (\citealt{Escudero22}), and NGC~1316, for which four populations were found (\citealt{Sesto16}). Moreover, some bimodal systems show a blue tilt in the distribution (\citealt{harris06}; \citealt{strader06}), in which the blue GCs become redder with increasing luminosities. Several hypotheses have been proposed to explain such behavior. These go from self-enrichment or the fact that these GCs possessed dark matter in the past (\citealt{strader06}), or by pre-enrichment of the more massive GCs from their larger parent gas clouds, which could hold more material (\citealt{harris06}). In this scenario, smaller GCs could form directly from pregalactic clouds, whereas the more massive ones form from bigger clouds that, due to the deeper potential well, could retain the gas after the first round of star formation, making the cluster self-enrich and gain higher metallicities (\citealt{harris06}). Moreover, E-MOSAICS simulations showed that the blue tilt effect appears to be stronger in more massive galaxies, and could be explained by the increase in GCs' mass with host galaxy metallicity in hierarchically assembling galaxies (\citealt{Usher}). \\
It has been shown that the relation between the number of GCs and the stellar mass of the host galaxy is nonlinear, with low-mass galaxies (especially dwarfs) and very massive galaxies more prone to produce GCs than stars with respect to galaxies of intermediate masses (\citealt{Beasley}). Moreover, the specific frequency of the GCSs depends on the galaxy’s morphology, with the highest values found in giant and dwarf ellipticals, whereas the lowest values belong to late-type galaxies (\citealt{harris}). It has been shown that the GCSs can extend from the inner regions of a galaxy to the outer haloes, with a halo population particularly numerous in early-type galaxies (\citealt{RC}).
Here, GCs can be observed up to $5-20~r_{\rm eff}$ (\citealt{RC2023}), whereas the 2D kinematics models of the stellar light can be obtained in detail for most galaxies inside $1~r_{\rm eff}$, and up to $2-4~r_{\rm eff}$ for some early-type galaxies (e.g., \citealt{Proctor09}; \citealt{Dolfi21}). Observations of stellar light at larger galactocentric distances are difficult to obtain due to the large telescope time required to probe different position angles (\citealt{Proctor09}), so the study of GCs is important to investigate the properties of the outer parts of the matter distribution of the host galaxy.
Finally, galaxies in clusters are likely to undergo multiple major and minor merger events. In fact, even though the velocity dispersion is large, the infall of galaxy groups provides a mechanism that promotes slow encounters and mergers within the cluster (\citealt{Mihos}). Moreover, simulations of tidal interactions in galaxy clusters show that, at the end of the simulation, massive galaxies are strongly clustered, and the probability of galactic collisions and mergers increases (\citealt{Gnedin}). In the case of minor mergers, all the material of the satellite galaxy is incorporated by the larger galaxy. These events are also likely to trigger and drive star formation in the case of early-type galaxies (\citealt{Kav}). Moreover, merger events can account for different populations of GCs, such as it was found in the low-mass early type galaxy NGC~4150 (\citealt{Kava}). In fact, models predict that the redder population of GCs would form in situ and would be more concentrated in the bulge of the host galaxy, whereas the blue population represents the accreted component and would reside in the halo (\citealt{Cote}). This was confirmed by E-MOSAICS simulations, where we can see that in situ and accreted GCs have comparable numbers in the inner halo, whereas GCs in the halo outskirts have a preferentially accreted origin (\citealt{RC}).

The Centaurus cluster is located at a distance of $42.5\pm 3.2$ Mpc (\citealt{mieske}). It is composed of the subcluster Cen~30, which represent the main component dominated by the giant elliptical NGC~4696 (see Tab.\ref{tab:galaxy}), and Cen~45, a spiral-rich subcomponent dominated by NGC~4709 (\citealt{Lucey}). The velocity distribution of a sample of cluster galaxies reveals the young, unrelaxed nature of the system, with Cen~45 being an infalling galaxy group (\citealt{Stein}). Interactions among galaxies are frequent in such an environment, and in the central region of Centaurus are confirmed by the presence of a filamentary structure connecting NGC~4696 to NGC~4696B, by a metallicity excess around NGC~4709 (\citealt{Walker}), and by asymmetric temperature variations in the X-ray gas with the hottest regions coinciding with NGC~4709 (\citealt{Churazov}). Besides, NGC~4696 is characterized by a filamentary structure crossing its central regions and extending towards NGC~4696B, suggesting that it underwent some merging phenomena in the past (\citealt{Churazov}) or that the material was acquired by ram-pressure stripping of NGC~4696B (\citealt{Walker}). Therefore, the identification and study of GC candidates around NGC~4696 has the potential to provide further hints about its assembly and evolution. In fact, disentangling between in situ or accreted formation of GCs could give important clues on the star formation history of this galaxy, the accretion phenomena, and its dark matter content. Furthermore, it is possible that the GCs host ultra- or hyperluminous X-ray sources, such as ESO~243-49~HLX-1 (\citealt{farrell}). In this case, the characterization of the X-ray sources population could give important information on the black holes and neutron star content, and on the merger history of NGC~4696.
Finally, giant elliptical galaxies are likely to show a large satellite galaxies population. Among these, the so called ultra-compact-dwarfs (UCDs) are particularly interesting since they show colors and metallicities between those of GCs and early-type dwarf galaxies (\citealt{Faifer17}), with masses between $10^6$ and $10^8$ M$_{\odot}$ (\citealt{hilker09}). Since the first discovery in 1999 (\citealt{Hilker99};\citealt{drink}), they have been detected in large numbers around external galaxies. Their origin is still unclear, but one of the most interesting hypothesis is that they could be the remnants of nucleated galaxies that were tidally stripped inside the cluster (\citealt{hilker09}). In this scenario, the study of UCDs, first identified around NGC~4696 by \citet{Mieske07}, could give us clues on the dynamical history of the galaxy.

In this paper, we present the identification and characterization of globular cluster candidates of NGC~4696, as well as a first catalog of GC candidates. The paper is organized as follows. In Sec.~\ref{sec:obs and data red} we describe the observations and data reduction, in Sec.~\ref{sec:GCsel} we present the selection criteria, in Sec.~\ref{sec:results} we present the analysis of the GC candidates, in Sec.~\ref{sec:color cut} we compare the results obtained using a different color cut, in Sec.~\ref{GCLF} we derive the GC luminosity function and specific frequency, and in Sec.~\ref{sec:conclusion} we discuss the results and provide the conclusions.

\section{Observations and data reduction}
\label{sec:obs and data red}
In this Section we describe the observation process (Sec.~\ref{sec:data}), the calculation of the model of the surface brightness distribution and of the residual images of NGC~4696 and of other five galaxies in the field (Sec.~\ref{model}), and the detection of the sources (Sec.~\ref{detection}).

\subsection{Description of the data}
\label{sec:data} % used for referring to this section from elsewhere

\begin{table}
\caption{Observational properties of the galaxy NGC~4696.} 
\centering
\resizebox{\columnwidth}{!}{\renewcommand{\arraystretch}{1.5}\begin{tabular}{c c c c c c c c c c c}
%\begin{tabular}{c c c c c c c c c c c}
\hline
    Name & RA & DEC & Type & $v_{\rm hel}$ & PA & $e$ & $d$ & $A_g$ & $A_r$ & $A_i$ \\
         & (J2000) & (J2000) & (RC3) & (km s$^{-1}$) & ($^{\circ}$) & & (Mpc) & (mag) & (mag) & (mag)  \\
    \hline
    NGC~4696 & $12^{\rm h}48^{\rm m}49^{\rm s}.8$ & $-41^{\circ}18'39''$ & cD pec & $2958$ & $95$ & $0.17$ & $42.5\pm 3.2$ & $0.368$ & $0.255$ & $0.189$ \\
\hline
\end{tabular}}
\tablefoot{The columns report the name of the object, the coordinates (RA and DEC), the type, the heliocentric velocity, and the position angle of its major axis (\citealt{deV}). The ellipticity was estimated by \citealt{Jarrett03}. The distance ($d$) was estimated by \citet{mieske} using the Surface Brightness Fluctuation method. The last three columns show the Galactic extinction coefficients in the three filters (\citealt{SF2011}).}
\label{tab:galaxy}
\end{table}

\begin{table}
\caption{Observations summary.}
     
    \label{tab:log}
    \centering
    \resizebox{\columnwidth}{!}{\renewcommand{\arraystretch}{1.5}\begin{tabular}{c c c c c c c}
    \hline
     Date & $T_{{\rm exp},g'}$ & $T_{{\rm exp},r'}$ & $T_{{\rm exp},i'}$ & FWHM$_{g'}$ & FWHM$_{r'}$ & FWHM$_{i'}$  \\
     $\rm{[dd/mm/yy]}$ & (s) & (s) & (s) & (arcsec) & (arcsec) & (arcsec) \\
    \hline
    $21-22/04/2018$ & $12300$ & $11100$ & $14700$ & $0.64$ & $0.6$ & $0.54$ \\
    \hline    
    \end{tabular}}
    \tablefoot{
     The columns report the date of the observation, the exposure times, and the FWHM of the PSF for the different filters.
     }

\end{table}

\begin{figure*}
    \sidecaption
    \includegraphics[width=12cm]{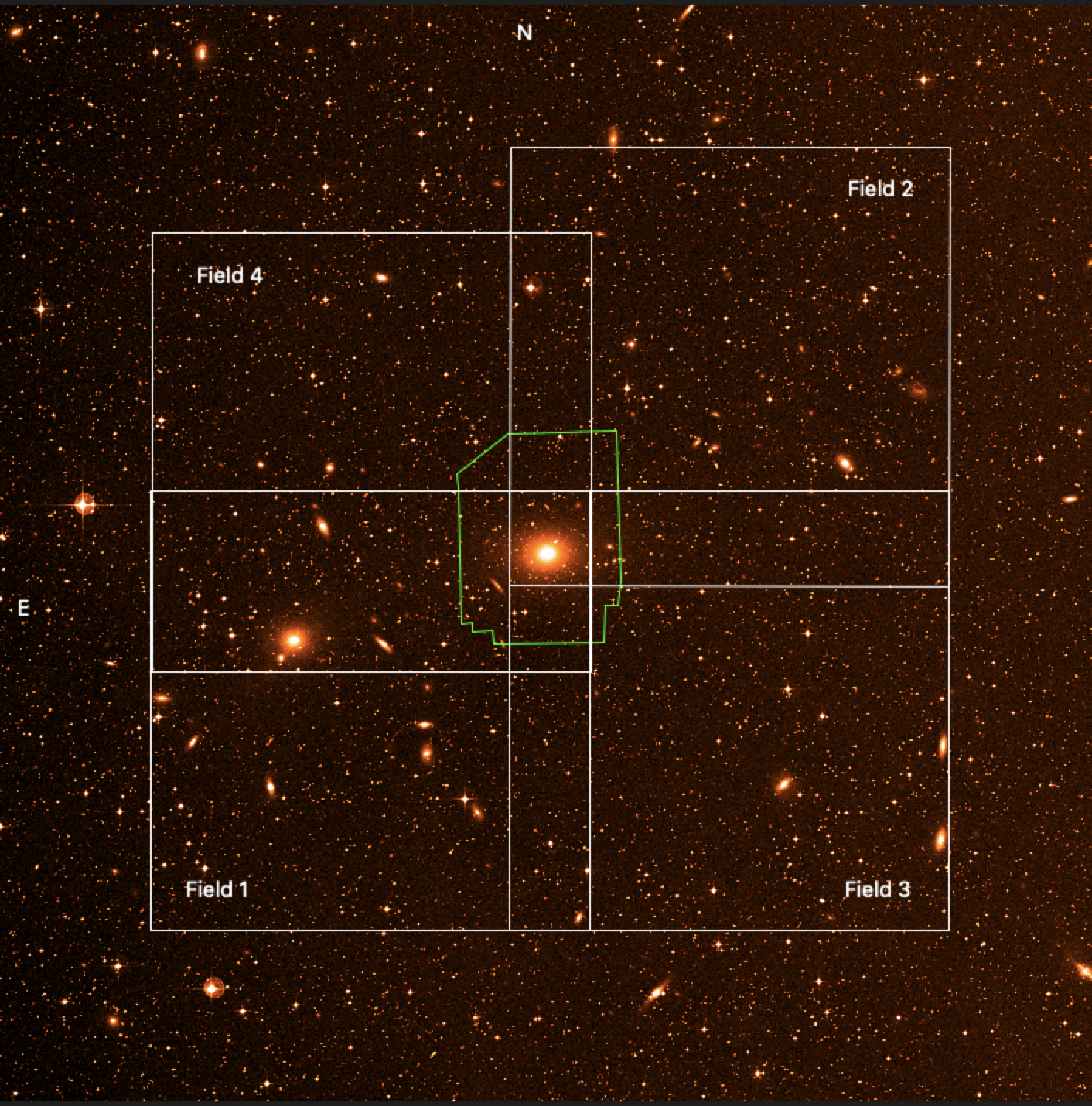}
    \caption{Centaurus cluster. The image shows the four fields obtained during the observation. The field-of-view is of $1\times 1$~deg$^2$, which corresponds to $0.742\times 0.742$~Mpc$^2$, with the north pointing in the upward direction and the east in the leftward direction. The green line delimits the region analyzed in this work.}
    \label{fig:Centaurus}
\end{figure*}

\begin{figure}
    \centering
    \includegraphics[scale=0.45]{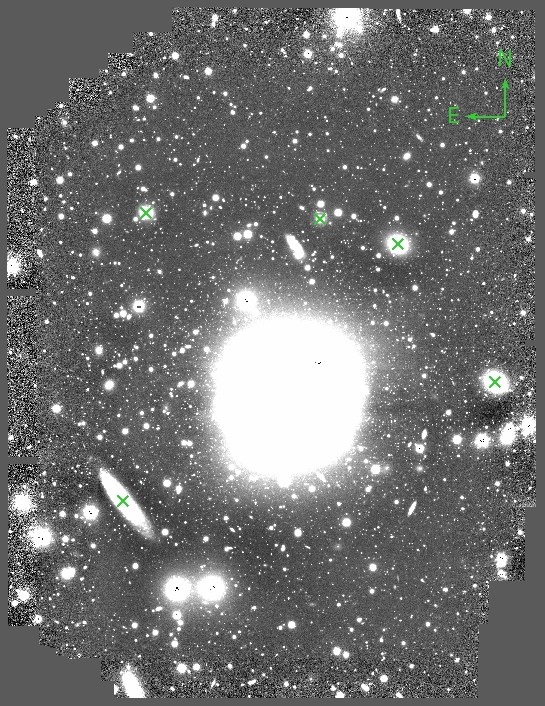}
    \caption{$g'$ band image centered on NGC~4696 obtained at the \textit{Magellan} telescope. The field-of-view (shown in green in Fig.~\ref{fig:Centaurus}) is of $\sim 8.795\times 11.494$ arcmin$^2$, which corresponds to $\sim 109\times 142$~kpc$^2$, with the north pointing the upward direction and the east in the leftward direction. The green crosses show the positions of the $5$ small companions of NGC~4696 for which we calculated the models of the surface brightness distribution (Tab.~\ref{tab:isofit properties}).}
    \label{fig:g}
\end{figure}

We observed four fields of the Centaurus cluster (Fig.~\ref{fig:Centaurus}) with MegaCam mounted on the \textit{Magellan} $6.5$ meter telescope during two nights in April 2018. MegaCam is a large mosaic CCD with a $24\times 24$ arcmin$^2$ field-of-view. The camera uses an array of $36$ E2V 42-90 backside-illuminated $2048\times 4608$~pixel CCDs, each covering a field-of-view of $160\times 360$~arcsec$^2$ and with a scale of $0.08$~arcsec/pixel. MegaCam is operated in campaign mode at the f/5 Cassegrain focus (\citealt{Megacam}). CCDs are mounted adjacent to one another, with a $6$~arcsec gap between them, thus a dithering pattern has to be followed to cover these gaps. The images were taken with a typical exposure time of $300$~s, in order to minimize saturation of bright sources (mostly foreground stars) and to minimize image degradation due to varying seeing conditions. In each of the four observed fields, NGC~4696 was placed close to the corner. The data presented here are the combination of the three adjacent CCDs that are common to all fields, with total exposure times in the innermost region common to all the four fields of $12300$, $11100$, and $14700$~s for the $g'$, $r'$, and $i'$ filters, respectively. The analysis is concentrated to a field-of-view of $8.795\times 11.494$~arcmin$^2$ around the giant elliptical galaxy NGC~4696 in the central region of the Centaurus cluster (Fig.~\ref{fig:g}), that include the region in common to the four fields shown in Fig.~\ref{fig:Centaurus}. In green, we overlay the region that was used for this analysis. We note that the total exposure time is not homogeneous across the indicated field, however exposure maps are used in our photometry to properly weight each region (see Fig.\ref{fig:exp}). The complex coverage arises from the random dithering pattern which was used to remove the physical gaps between the CCDs composing MegaCam. The observation log is summarized in Tab.~\ref{tab:log}. 
Most of the data reduction was performed through the dedicated MMT/\textit{Magellan} MegaCam pipeline (\citealt{Megacam15}). Specifically, tasks like WCS fitting, fringe and illumination corrections were all done automatically and carefully checked at each reduction step. The final user-ready images produced by the pipeline are the individual multi-extension fits files as well as a preliminary mosaic of the entire field. The stacked image produced by the pipeline has been optimized for the detection of compact sources. Therefore, we prefer to use our own set of parameters in order to keep large scale structures that could otherwise be partly filtered. While the goal of this work is related to compact stellar systems, in a forthcoming study (Federle et al., in prep.) we will analyze the intra-cluster light and other faint structures that need a finer tuning of the mosaicing parameters. For this work, we use the individual fits files as processed by the pipeline, but with our own parameters for the stacked mosaic, which was done with Swarp (\citealt{swarp}) and using Astropy (\citealt{astropy1}; \citealt{astropy2}) to perform preliminary statistics on common empty sky regions. The quality and photometric homogeneity of the stacked images has been carefully checked in each step. We also noticed that the pipeline was in some cases not enough to eliminate the fringe pattern in the $i'$ filter. We therefore applied a custom fringe pattern taken from different fields, properly scaled to the exposure times in this band. After this procedure, no systematic residuals could be observed in any of the stacked images. We present here the analysis of the central $\sim 8.795\times 11.494$~arcmin$^2$ of the Centaurus cluster, with particular focus on the giant elliptical NGC~4696. The final $g'$ image of this field is shown in Fig.~\ref{fig:g}.

\subsection{Subtraction of the galaxy light distribution}
\label{model}
In order to maximize the detection and perform photometry of the GC candidates, we first calculated the models of the surface brightness distribution of NGC~4696 and of other five galaxies in its proximity for each filter and subtracted them from the images. In this way the residual images show compact sources otherwise immersed in the galaxies light. The models of the surface brightness distribution were obtained using {\tt ISOFIT} and {\tt CMODEL} packages (\citealt{ciambur}) in IRAF\footnote{IRAF is distributed by the National Optical Astronomy Observatory (NOAO), directed by the Association of University for Research in Astronomy (AURA) together with the National Science Foundation (NSF). It is available in http://iraf.noao.edu/}. These are implementations of the {\tt ELLIPSE} and {\tt BMODEL} tasks (\citealt{jed}), which allow to calculate the model of the surface brightness distribution by fitting elliptical isophotes to the galaxy via a Fourier series expansion. The surface brightness of the isophotes is calculated as:\begin{equation}
    I(\psi)=\langle I_{\rm ell}\rangle + \sum_{n=1}^{\infty}A_n\sin{(n\psi)} + \sum_{n=1}^{\infty}B_n\cos{(n\psi)}
\end{equation}
where $\langle I_{\rm ell}\rangle$ is the surface brightness the isophote would have if it was a perfect ellipse, $\psi$ is the eccentric anomaly, and $A_n$ and $B_n$ are the Fourier coefficients (\citealt{ciambur}).

We used {\tt ISOFIT} because we find it to be more accurate than {\tt ELLIPSE} in the interpolation of the isophotes in this case. In fact, it samples them using the eccentric anomaly, which results in equal length arcs on the ellipse and describes better the deviations from the elliptical form of the isophotes. In order to calculate these deviations, the program allows the user to add higher harmonics and use them as input for the {\tt CMODEL} task. For the case of NGC~4696, we used the first four harmonics, which are given by default and describe the coordinates of the center of the galaxy, the ellipticity and the position angle, and we add the $n=6$ and $n=8$ harmonics in the $g'$ and $i'$ bands, and the $n=6$ harmonic in the $r'$ band. We only used the even harmonics because the odd ones introduce a negligible correction in the case of a galaxy with no significant asymmetries (\citealt{ciambur}), such as NGC~4696. The models were calculated after masking external galaxies and point-like sources present in the field, in order to have a precise measure of the light of the galaxy under consideration. The properties of the other five galaxies analyzed, and the parameters used in {\tt ISOFIT} are listed in Tab.~\ref{tab:isofit properties} .

\begin{table*}
\caption{Properties of the five galaxies for which the models of the surface brightness distribution were calculated.}
    \centering
    \begin{tabular}{c c c c c c c}
    \hline
    Name & RA & DEC & Type & $v_{\rm hel}$ & $d$ & $n$ \\
     & (J2000) & (J2000) & (RC3) & (km~s$^{-1}$) & (Mpc) & \\ 
    \hline
    WISEA J124839.70-411605.1 & $12^{\rm h}48^{\rm m}39^{\rm s}.69$ & $-41^{\circ}16'05".60$ & dE & $2851.03$ & $46.21\pm 3.26$ & $6-8-10$ \\
    WISEA J124831.02-411823.2 & $12^{\rm h}48^{\rm m}31^{\rm s}.03$ & $-41^{\circ}18'23".19$ & dE & $3003.02$ & $48.46\pm 3.41$ & $6-8-10$ \\
    WISEA J124902.02-411533.7 & $12^{\rm h}49^{\rm m}02^{\rm s}.00$ & $-41^{\circ}15'33".30$ & dE & $2074.86$ & $34.76\pm 2.66$ & $6$  \\
    WISEA J124846.61-411539.4 & $12^{\rm h}48^{\rm m}46^{\rm s}.61$ & $-41^{\circ}15'39".50$ & ? & ? & ? & $6-8-10$ \\
    ESO 322- G 093 & $12^{\rm h}49^{\rm m}04^{\rm s}.12$ & $-41^{\circ}20'19".60$ & Sc & $3468.90$ & $55.32\pm 3.89$ & $6-8-10$ \\
    \hline
    \end{tabular}
    \tablefoot{The columns show the name of the galaxy, the coordinates (RA and DEC), the type, the heliocentric velocity, and the distance as reported in NED\protect\tablefootnote{https://ned.ipac.caltech.edu/}. The last column show the numbers of the higher order harmonics used in {\tt ISOFIT} to calculate the models of the surface brightness distribution, which were equal for the three filters.}
    \label{tab:isofit properties}
\end{table*}

\subsection{Source detection and photometry}
\label{detection}
After the subtraction of the models from the images, we run the {\tt SExtractor} package (version 2.19.5; \citealt{sex}) to detect sources. After some experimentation, we chose a Gaussian filter, a detection threshold of $3\times \sigma_{\rm sky}$, a detection minimum area of $5$~pixels, a background filter size of $3$, and different background meshes (ranges from $4$ to $24$) in order to maximize the detection of point-like sources. We found $14507$ sources in common in the three filters. The magnitudes of the sources were calculated using the parameter {\tt MAG\_ISO} of {\tt SExtractor}, which in our case has typically lower errors with respect to {\tt MAG\_APER}. Moreover, since we noticed that in both \citet{harris06} and  \citet{Harris23} the color distribution of the GC candidates show a deep between the blue and red populations that is not evident in our analysis, we compared the color distribution obtained with {\tt MAG\_ISO} with that obtained with {\tt MAG\_APER}. However, neither the use of {\tt MAG\_APER} shows the deep between the red and blue populations in the color distribution, so for the following analysis we use {\tt MAG\_ISO}. The photometric calibration was performed using standard stars from the NOAO Source Catalog (\citealt{nsc}). We found $3490$ sources in common between ours and the NSC catalogs. For the calibration we considered only those sources for which in the NSC catalog the number of observations was $>1$, and that have a stellarity index $>0.5$ in order to exclude extended sources. Moreover, we visually checked the remaining $925$ objects in order to exclude the ones that were located near other sources or that were too immersed in the light of the galaxy. After the inspection, we decided to use $105$ sources to calibrate our sample. The rms scatter of the observed minus catalog magnitudes is of $0.02$~mag. Finally, we corrected the magnitudes for Galactic extinction according to the values shown in Tab.~\ref{tab:galaxy}. The magnitudes reported in this work are in the AB system.

\begin{figure}
    \centering
    \includegraphics[scale=0.3]{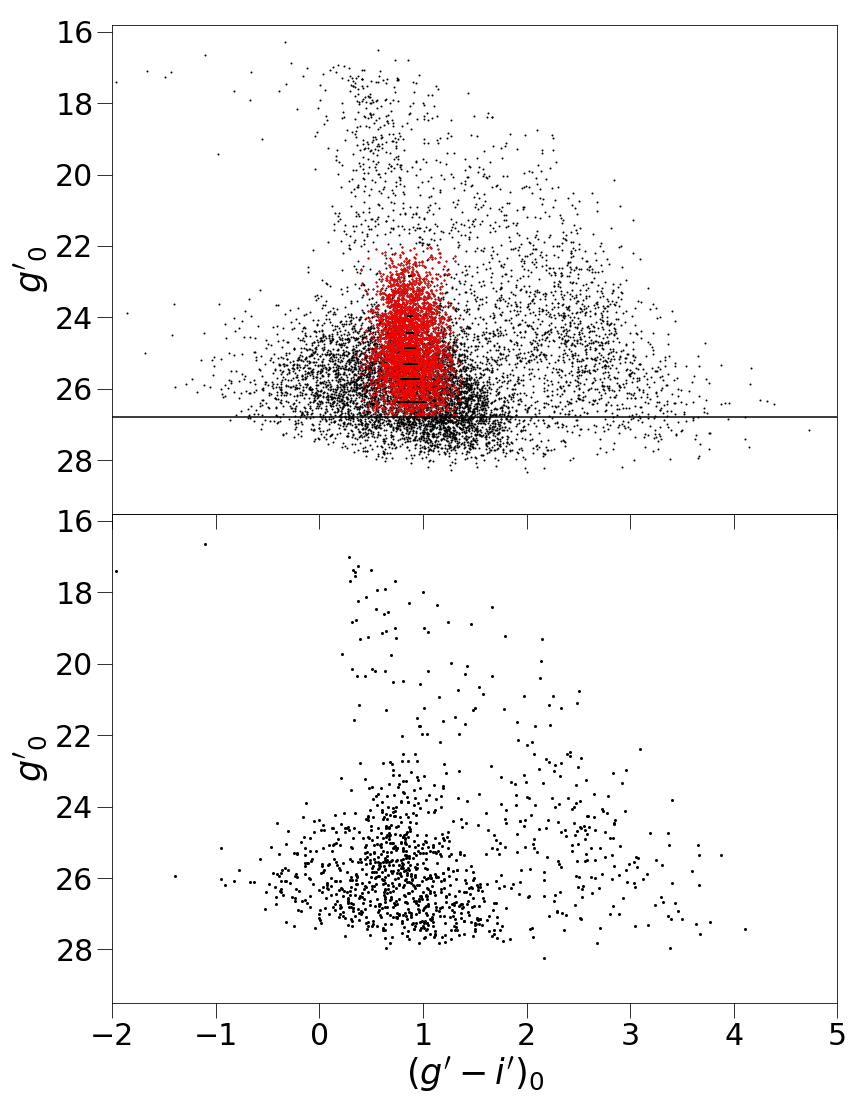}
    \caption{Comparison between the color magnitude diagram of the sources and that of the background regions. {\em Top panel}: Color-magnitude diagram of the $10970$ sources (black dots) selected according to their shape. The plot shows the $g'_0$ magnitude as a function of the $(g'-i')_0$ color. The horizontal line at $g'_0=26.789$ shows the limit applied as conservative measure in order to avoid eventual color trends due to the incompleteness of the sample. The red dots represent the sources after the selection on color and magnitudes, and the black bars are the mean errorbars in the magnitude bins used to calculate the color distributions and the positions of the blue and red peaks. {\em Bottom panel}: Color-magnitude diagram of the background regions. The plot shows the $g'_0$ magnitude as a function of the $(g'-i')_0$ color for the $1138$ sources located in the two rectangle regions of $6\times 1$~arcmin$^2$ used to calculate the background level.}
    \label{fig:cmd_class}
\end{figure}

Besides the magnitude of the sources, {\tt SExtractor} allows the user to estimate the type of object analyzed with the {\tt CLASS\_STAR} parameter, which has values between $0$ and $1$ (\citealt{sex}). In particular, objects with ${\tt CLASS\_STAR}=1$ are stellar objects, while those with ${\tt CLASS\_STAR}=0$ are extended objects. In order to separate GCs from galaxies, we divided the detected sources into two subsamples with ${\tt CLASS\_STAR}>0.5$,  and ${\tt CLASS\_STAR}\leq 0.5$. The typical effective radius of GCs is $r_{\rm eff}\sim 3$~pc, which corresponds to $\sim 0.015$~arcsec at the distance of Centaurus. This means that most of them are unresolved, so we selected sources with ${\tt CLASS\_STAR}>0.5$, in the $i'$ filter, which corresponds to the image with the best seeing conditions. In this way, we found $10970$ compact (or unresolved) sources, which are shown in the color-magnitude diagram in the top panel of Fig.~\ref{fig:cmd_class}. 

\section{Selection of globular cluster candidates}
\label{sec:GCsel}
\begin{figure}
    \centering
    \includegraphics[scale=0.32]{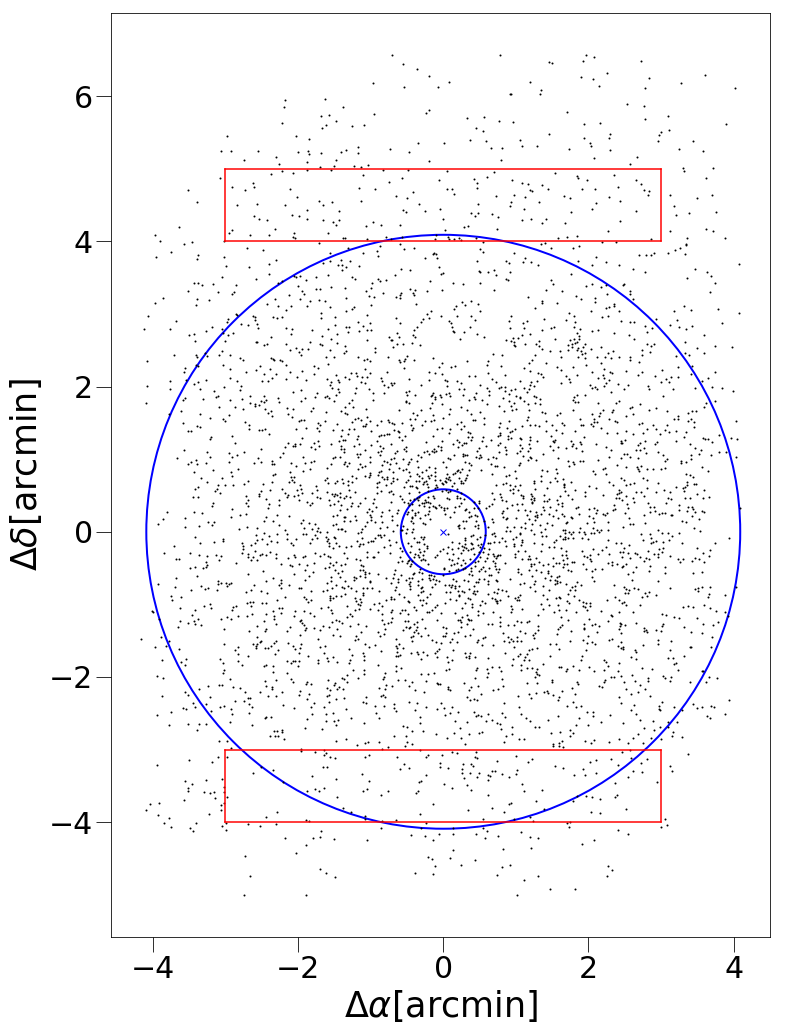}
    \caption{Positions of the globular cluster candidates with respect to the center of the galaxy. The plot shows declination and right ascension of the sources, in which the center of NGC~4696 is located at ($\Delta \alpha , \Delta \delta)=(0,0)$. The blue circles define the borders of the annulus inside of which the azimuthal distribution of the GCs was calculated, whereas the red rectangles represent the regions where the contamination level from background objects and foreground stars was calculated.}
    \label{fig:positions}
\end{figure}

Fig.~\ref{fig:cmd_class} shows a very large spread in color of the selected sources. In order to eliminate potential contaminators from the $10970$ compact sources sample, such as foreground and background sources, we applied a second selection criterion based on $(g'-r')_0$ and $(r'-i')_0$ colors, and on the $g'_0$ magnitudes. In particular, we used $0.35<(g'-r')_0<0.85$, $0.0<(r'-i')_0<0.55$, and $g'_0>22.0$ as described in \citet{Faifer17}. These limits were obtained in \citet{Faifer11} for the GCSs of five early type galaxies located at a distance modulus between $29.93\pm 0.09$ and $31.90\pm 0.20$. Given these distances and that of NGC~4696, we can apply the same color selection since redshift effects are negligible. The color limits were obtained by \citet{Faifer11} by considering the completeness of the GC sample and the magnitudes of Galactic globular clusters. In particular, the bright limit corresponds to $M_I>-12$~mag, which represents the typical separation between GCs and UCDs. Moreover, as shown in Fig.~8 in \citet{Faifer11}, the limits cover the metallicity range expected for GCs. Our selection considers a somewhat larger range in color to take into account for the increasing errors going toward fainter magnitudes. Using these limits, we obtained a sample of $3973$ GC candidates, whose properties are summarized in Tab.~\ref{tab:catalog}. From Fig.~\ref{fig:cmd_class}, we can see some black dots in the same region of our selected sources. This is given by the fact that the color selection was made only on the $(g'-r')_0$ and $(r'-i')_0$ colors. Fig.~\ref{fig:positions} shows the positions of the sources with respect to the center of the host galaxy.

\section{Results}
\label{sec:results}

\subsection{Color-magnitude and color-color diagrams}
\label{sec:cmd}

\begin{figure}
    \centering
    \includegraphics[scale=0.25]{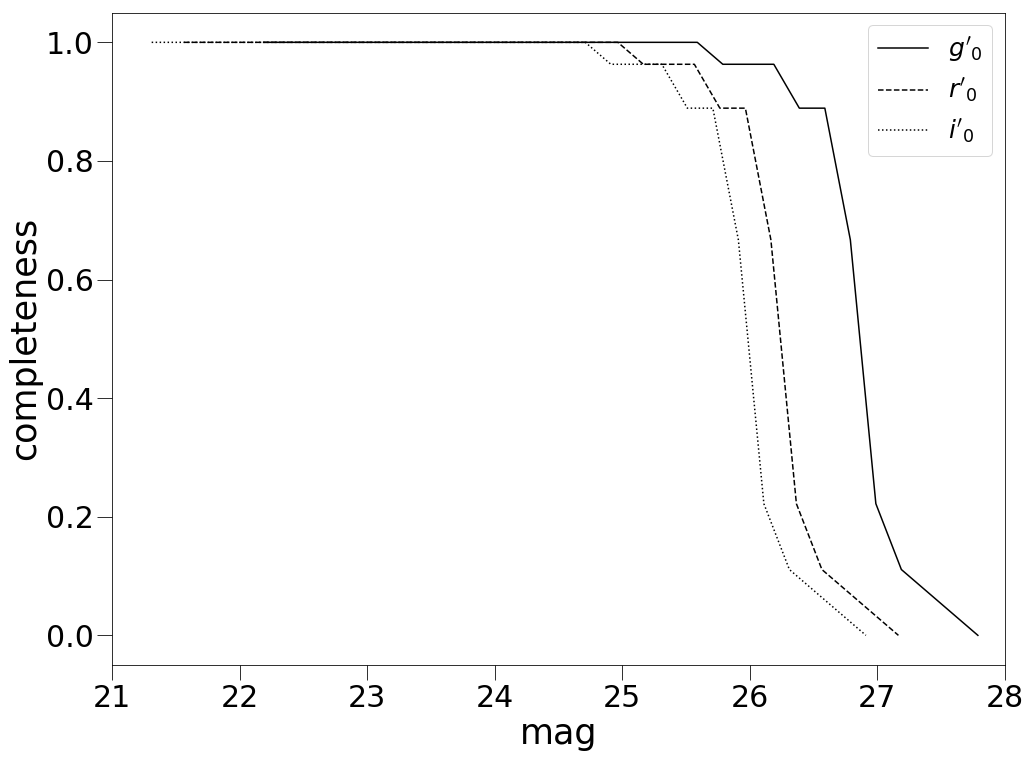}
    \caption{Completeness test results. The solid, dashed and dotted lines show the completeness for the $g'$, $r'$ and $i'$ bands respectively.}
    \label{fig:completeness}
\end{figure}

\begin{figure}
    \centering
\includegraphics[scale=0.25]{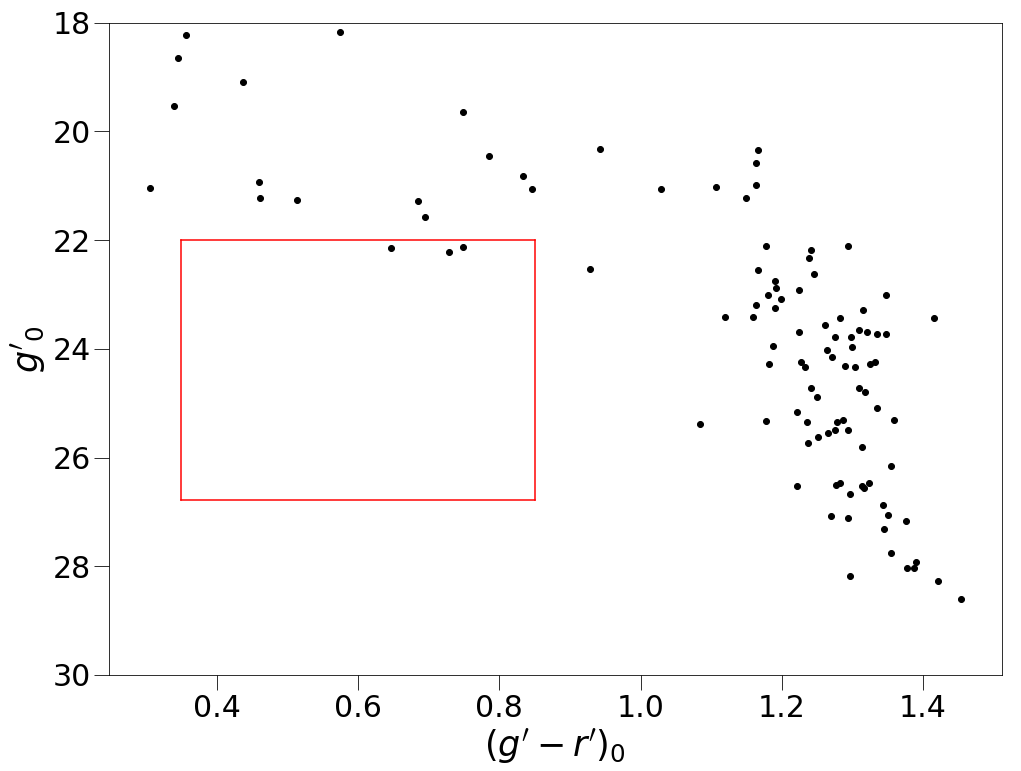}
    \caption{Color-magnitude diagram of the sources in the Besan{\c{c}}on model simulation. The plot shows the $g'_0$ magnitude as a function of the $(g'-r')_0$ color, for the sources in an area equal to that used for our background fields ($6\times 1$~arcmin$^2$). The red square shows the limits given by our GCs selection criteria.} 
    \label{fig:besancon}
\end{figure}

At this point, the sample of GC candidates, defined using a selection on the stellarity index (${\tt CLASS\_STAR}>0.5$ in the $i'$ filter), the colors ($0.35<(g'-r')_0<0.85$ and $0.0<(r'-i')_0<0.55$) and magnitude ($g'_0>22$), includes $3973$ sources. In the color-magnitude diagram (Fig.~\ref{fig:cmd_class}) we note that a color dependence of the limiting magnitude can be spotted as the mild trend at fainter magnitudes: blue GC candidates are more prone to be detected than red ones, within the color range defined in Sec.~\ref{sec:GCsel}.  
In order to test the detection limits of our photometry, we performed the completeness test of the GC sample. In particular, we used the {\tt ADDSTAR} task (\citealt{Stetson}) in {\tt IRAF} to add $200$ artificial star-like objects per each magnitude considered based on the PSF estimated with {\tt DAOPHOT}. The task allows the user to add point-like sources at chosen magnitude and positions, which in our case is important to recover their colors. The artificial stars were then added in the three filters at chosen positions in steps of $0.2$ magnitudes, so the total number of sources added with magnitude of $16.2\leq g'_0\leq 29.0$~mag is $12000$. The addition of $200$ objects in each step allows to get a good number statistics, but also to avoid crowding effects. Using the same {\tt SExtractor} parameters as described before, we determined how many of the added stars could be recovered in the three filters. The magnitudes were then corrected according to the same zeropoint and extinction values as in the precedent analysis (Tab.~\ref{tab:galaxy}). Moreover, we made sure that the artificial stars were added in the same positions in all the three filters, and that they had a color of $(g'-i')_0=0.878$ which represent the mean value of our sample. As can be seen in Fig.~\ref{fig:completeness}, the $50\%$ completeness limit is reached at a magnitude $g'_0=26.889$~mag, so by choosing to limit our study at $g'_0\leq 26.79$~mag, where we have a $67\%$ completeness, we are well above the $50\%$ completeness.

Despite the morphological and color selection criteria, some of the point-like sources are likely to be foreground stars or background galaxies, so an additional correction must be applied. For an upper limit estimate of the potential background contamination, we selected two rectangular regions of $6\times 1$~arcmin$^2$, located in the upper and lower parts of the field of view (see Fig.~\ref{fig:positions}), where they in part overlap the circular region because the image is not fully symmetric around NGC~4696. The radius of the circular region is of $4.091$~arcmin ($2.4~r_{\rm eff}$), and it was chosen in order to have the largest possible symmetrical region around NGC~4696 within the observed field. We run the photometry and the completeness test on the rectangular regions using the same parameters as before. The total number of sources in the background regions is of $1138$, whereas the number of sources with $g'_0>22$~mag is of $526$ and $530$ for the upper and lower regions, respectively. The color-magnitude diagram of these sources is shown in Fig.~\ref{fig:cmd_class}. As can be seen, it looks much like the CMD of the central one, supporting the hypothesis that it is also dominated by GCs. Of the sources added with the task {\tt ADDSTAR}, $16$ fall inside the upper region and $15$ in the lower one. 

Finally, to estimate which fraction of sources in the above mentioned background fields corresponds to foreground stars in the Milky Way, we compared our results with the star counts given by the Besan{\c{c}}on model of the Galaxy\footnote{Available in: https://model.obs-besancon.fr} (\citealt{Robin}). We run the simulation for the MegaCam filters, considering a region with the same area as our background regions and with $18.0\leq g'_0\leq 28.0$~mag. With these values, we found $104$ sources, of which only $3$ are within the range of the color-magnitude cut applied to select our GC candidates (Fig.~\ref{fig:besancon}). From this result, we can conclude that the contribution of foreground stars to the background level is negligible. In order to have an estimation of the contribution from background galaxies we would need to go to much larger galactocentric distances. For the time being, we will then assume the background level to be negligible.

\begin{figure}
    \centering
    \includegraphics[scale=0.25]{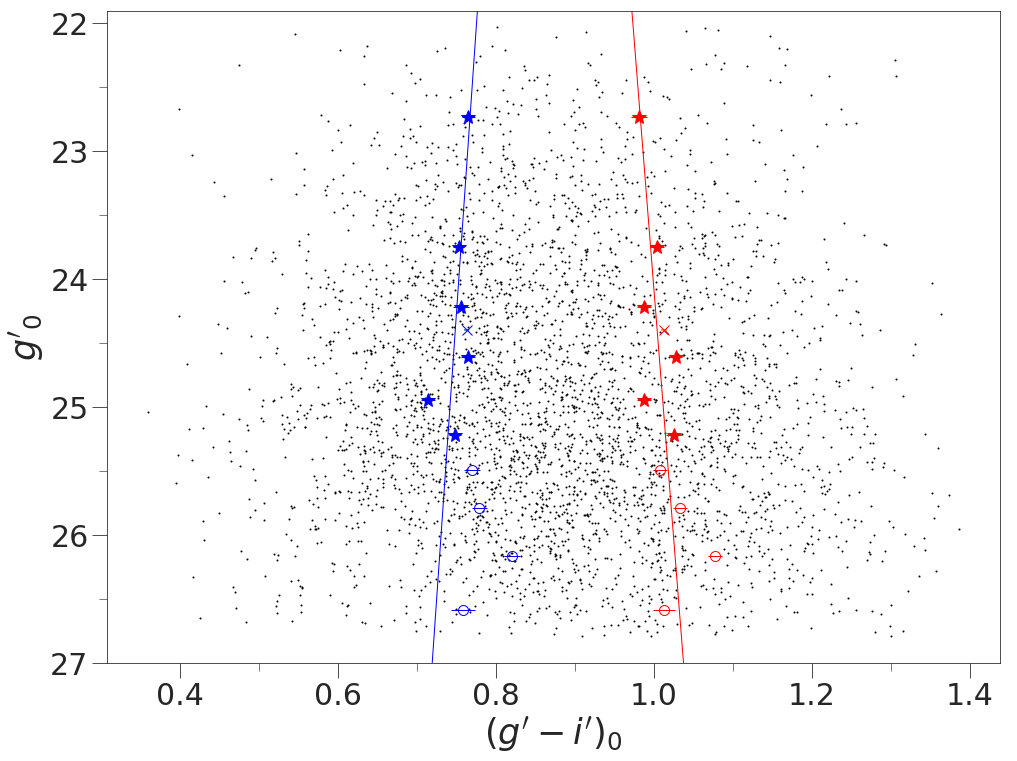}
    \caption{Color-magnitude diagram of the $3818$ globular cluster candidates. The plot shows the $g'_0$ magnitude as a function of the $(g'-i')_0$ color. The blue and red points correspond to the peaks of the blue and red GC populations, obtained by forcing the fit of two Gaussians on each subsamples of GCs. The blue and red star symbols represent the bins for which the bimodal distribution was confirmed by the Akaike Information Criterion. The blue and red x symbols represent the peaks of the total population of GC candidates, whereas the circles represents the bins for which the Akaike Information Criterion gave a non-bimodal distribution. The blue and red lines correspond to the best linear fits for the peaks in the color distributions.}
    \label{fig:cmd 3894}
\end{figure}

\begin{figure*}
    \sidecaption
    \includegraphics[width=12cm]{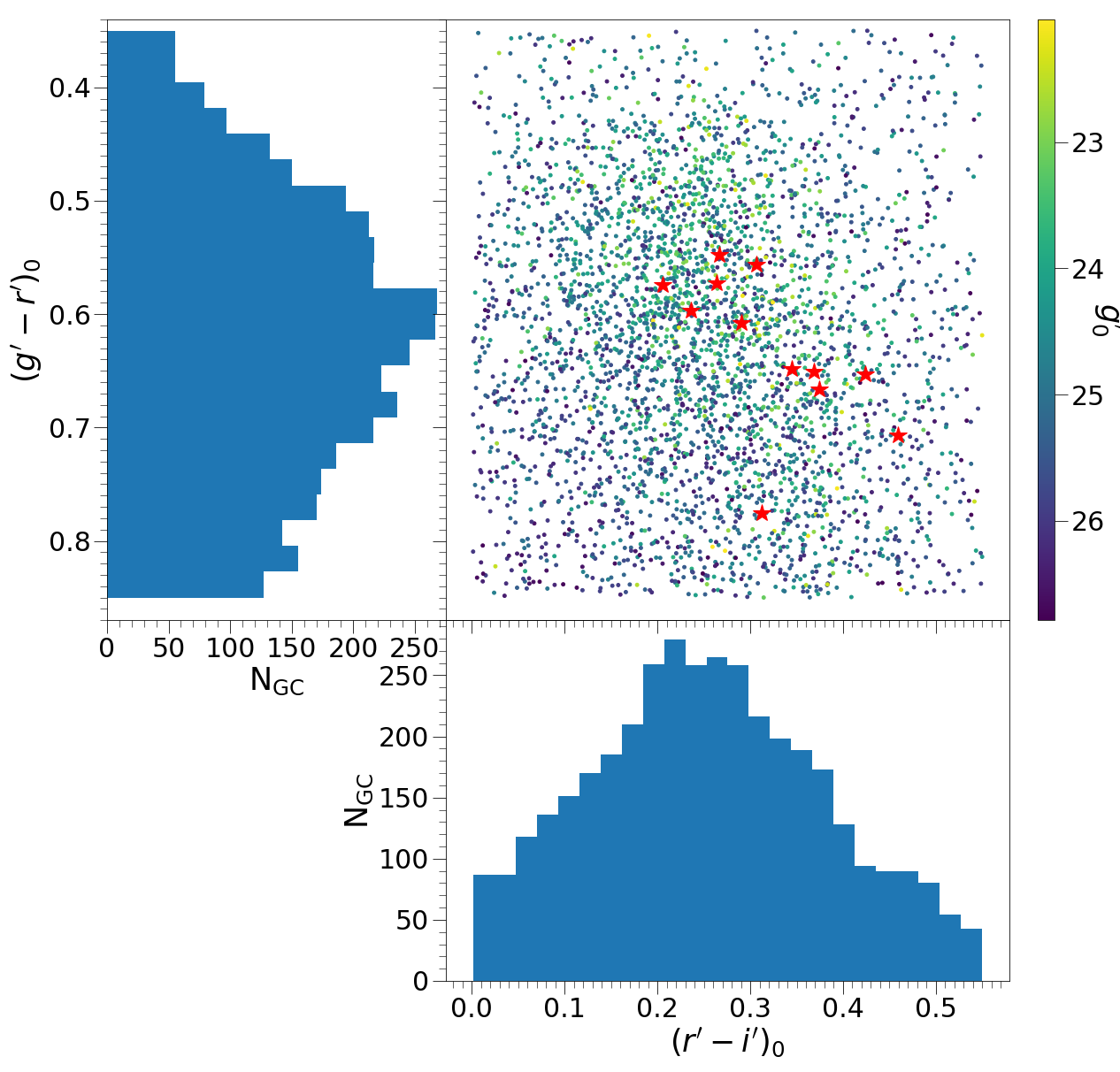}
    \caption{x versus y color-color diagram of the $3818$ globular cluster candidates selected according to their shape, color, and magnitude. The plot shows the $(g'-r')_0$ as a function of $(r'-i')_0$, color coded according to the $g'_0$ magnitude. The left and bottom panels show the $(g'-r')_0$ and $(r'-i')_0$ color distributions, respectively. The red stars represent compact objects identified by \citet{Mieske07} which were recovered in our selection. The properties of these objects are summarized in Tab.~\ref{tab:mieske comparison}.}
    \label{fig:colorcolor}
\end{figure*}

With the faint magnitude cut given by the completeness test described above, the number of GC candidates, shown by the red dots in Fig.~\ref{fig:cmd_class}, reduces to $3818$. From the photometry of the GC candidates we obtained the color-magnitude diagram, and the color-color diagram and color distributions, which are shown in Fig.~\ref{fig:cmd 3894} and Fig.~\ref{fig:colorcolor} respectively. We compared our catalog of point-like sources with the $27$ compact objects, for which a membership to the Centaurus cluster was spectroscopically confirmed by \citet{Mieske07}. We found $17$ sources in common, of which $12$ belong to our sample of $3818$ GC candidates, whereas the remaining $5$ sources belong to the sample after the selection made only on the stellarity index. For comparison, Fig.~\ref{fig:colorcolor} shows the $12$ compact sources belonging to our GC candidates sample. As can be seen, the color properties of these compact objects are well described within our selection criteria.

\subsection{Color distribution}
\label{sec:color distr}

\begin{figure*}
    \centering
    \includegraphics[width=\textwidth]{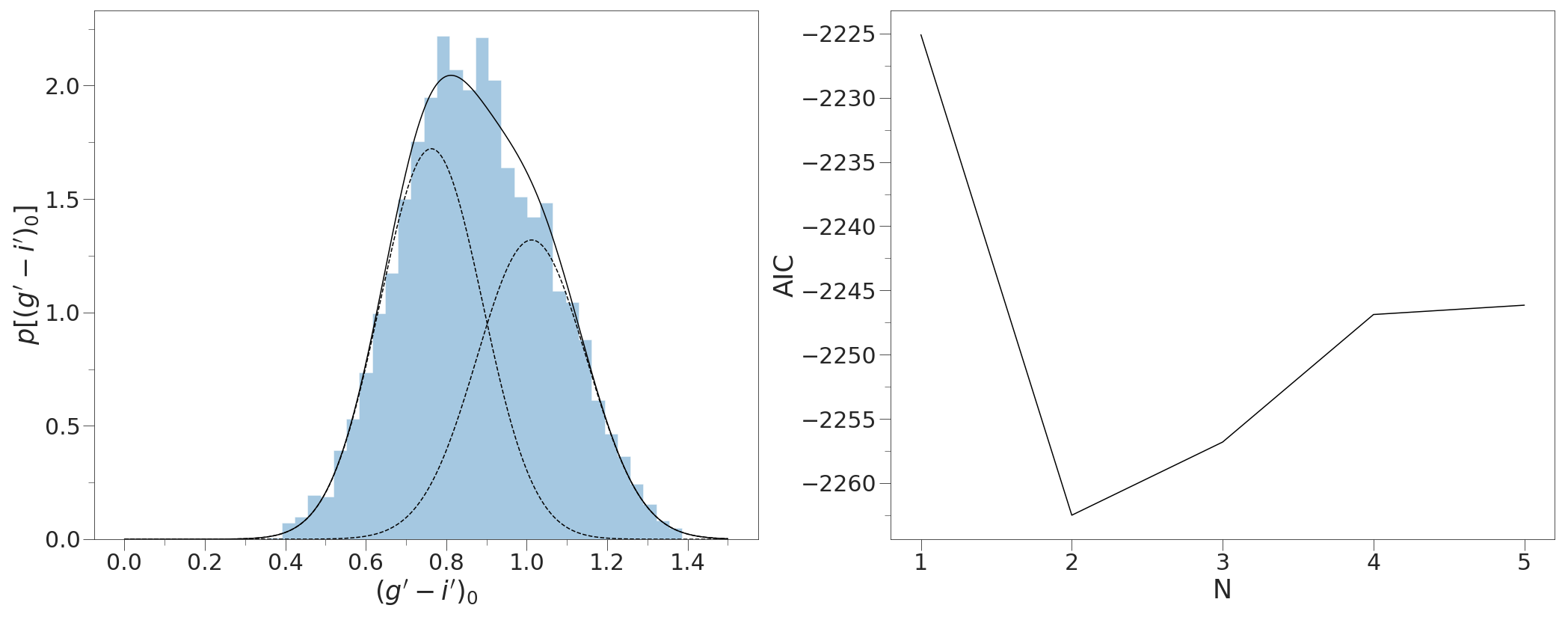}
    \caption{Akaike Information Criterion test. {\em Left panel}: probability density function vs the $(g'-i')_0$ color. The black line represents the best fit model, the dashed lines represent the two Gaussians described by the model. The color distribution is divided in bins according to the Freedman-Diaconis rule (\citealt{Freedman}). {\em Right panel}: values of the Akaike Information Criterion (AIC) as a function of the number of components in the model.}
    \label{fig:akaike}
\end{figure*}

As indicated by Fig.~\ref{fig:cmd 3894}, the $(g'-i')_0$ color distribution is clearly more complex than a single Gaussian. Bimodality has been found in most GCSs of giant galaxies and indeed NGC~4696 shows a distribution that could be well represented, at first glance, by two Gaussians.
In order to investigate this issue, we performed the analysis of the color distribution of the GC candidates with the Gaussian Mixture Modeling of {\tt sklearn} in Python (\citealt{Ped}), which allows to fit multiple Gaussians to check for multimodality. From this, we find that a bimodal Gaussian,  with the blue peak at $(g'-i')_0=0.763\pm 0.004$~mag and the red peak at $(g'-i')_0=1.012\pm 0.004$~mag, is preferred over a unimodal and a trimodal one, see right panel of Fig.~\ref{fig:akaike}. The red and blue populations were divided at the color for which the GMM fit gave an equal probability for the object of belonging to the red or blue subsamples (as described in \citealt{Escudero}). In our case the color cut of the total sample corresponds to $(g'-i')_0=0.905$ mag, which, given the transformations from Tab.~3 of \citet{Jordi06}, is bluer than the value found by \citet{harris06}. To test whether two Gaussians are preferable instead of three or more Gaussians, we used the Akaike Information Criterion (AIC; \citealt{Akaike}) implemented in astroML (\citealt{astroML}), which, given the data and a certain number of components for the model, estimates the quality of the model. We performed the AIC test using a number of components from $1$ to $5$. As can be seen in Fig.~\ref{fig:akaike} the best fit, for which the value of the AIC is minimum, is obtained for a $2$ Gaussian model.

\subsection{The blue tilt}
\begin{figure*}[h!]
    \sidecaption
    \includegraphics[width=12cm]{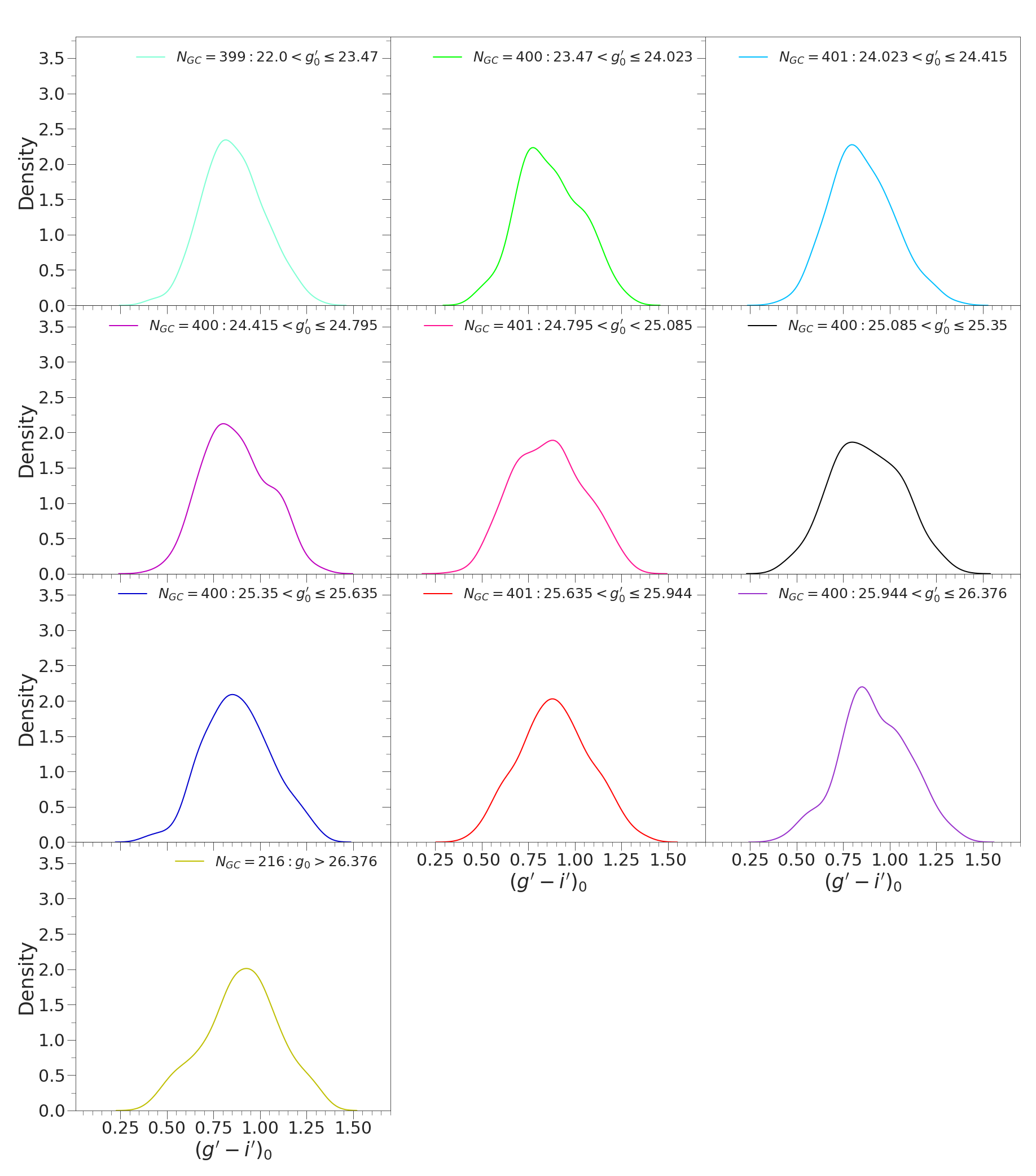}
    \caption{Color distribution of the subsamples of globular clusters. The plots go from the brightest magnitude bin (top left) to the faintest magnitude bin (bottom). The legends indicate the number of GCs for the considered bin. The number of components that better fit the color distributions in each bin according to the Akaike Information Criterion are summarized in Tab.~\ref{tab:GMM}.}
    \label{fig:col dist all}
\end{figure*}

\begin{figure}
    \centering
    \includegraphics[scale=0.2]{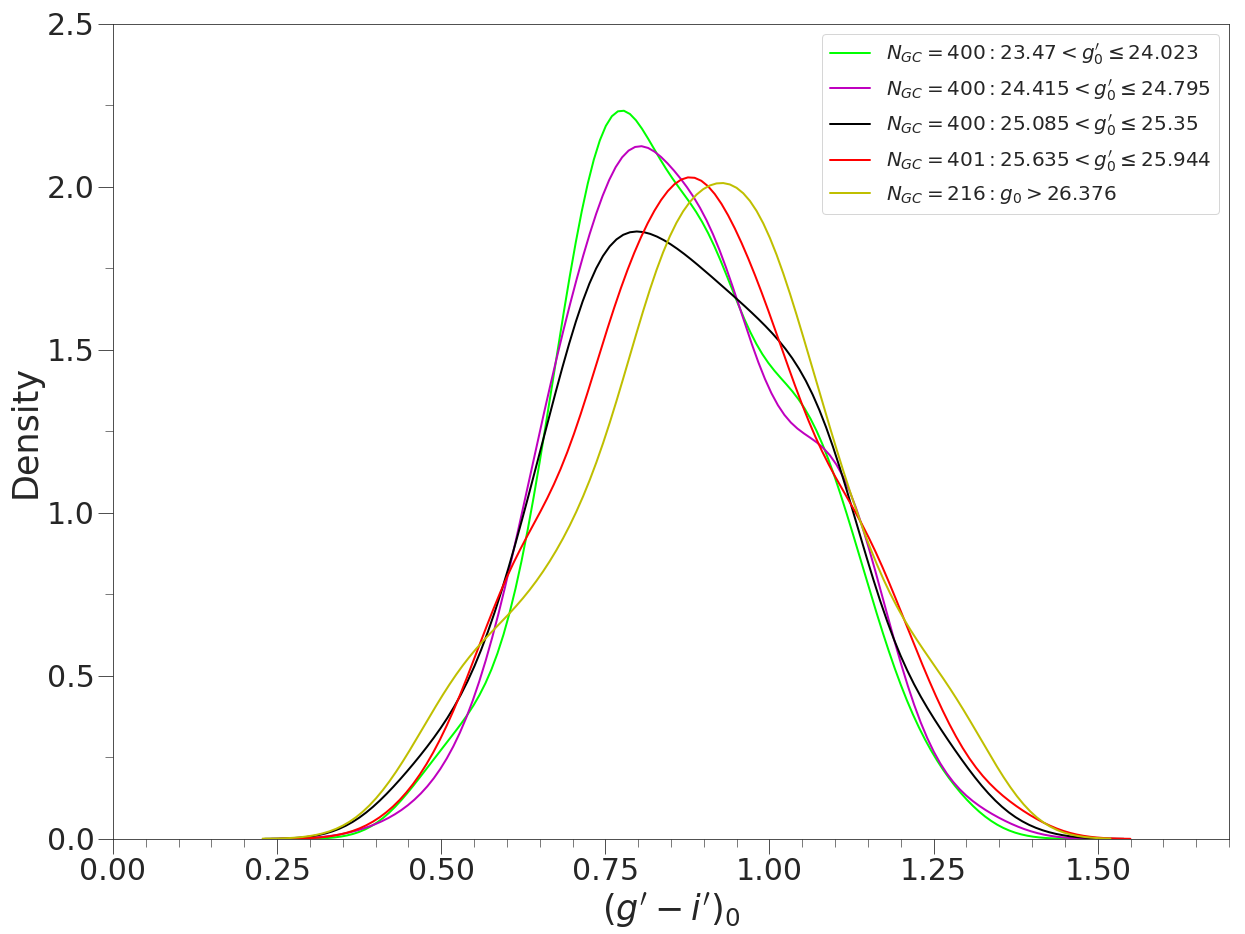}
    \caption{Color distribution of the subsamples of GCs. Each curve shows the color distribution within the indicated magnitude ranges. The widths of the magnitude bins were chosen to have a comparable number of sources per range.}
    \label{fig:total color}
\end{figure}

A large fraction of GCSs around external galaxies shows the so-called blue tilt in the color distribution (e.g., \citealt{Forbes10}; \citealt{Faifer11}; \citealt{harris06}; \citealt{mieske}). This means that, in a bimodal color distribution, the blue peaks become redder for increasing cluster's luminosities. In order to investigate this, we divided the GC candidates in subsamples according to their magnitudes, and making sure to have a comparable number of candidates in each bin in order to have a good number statistic. We then calculated the color distribution for each bin using the Kernel Density Estimation (KDE). This is a nonparametric method, which allows to estimate the shape of a density function, given a sample from the distribution (\citealt{KDE}). For our analysis, we calculated the KDE using a Gaussian kernel, whose shape and size was kept fixed for each magnitude bin. Fig.~\ref{fig:col dist all} shows the KDE of the color distribution for each subsample with the respective number of GCs, while Fig.~\ref{fig:total color} shows the superimposition of the color distributions. As can be seen from the plots, in some of the bins the color distributions are not unimodal, but they show at least two peaks. This was confirmed by the results of the Akaike Information Criterion summarized in Tab.~\ref{tab:GMM}. In particular, the positions of the blue and red peaks are shown in Fig.~\ref{fig:total color}. However, considering the errorbars of the linear fits we can not detect a significant blue tilt Fig.~\ref{fig:cmd 3894}. In fact, the slope of the linear fits is consistent with zero for both the blue:\begin{equation}
    (g'-i')_0=-0.014(\pm 0.018)*g'_0+1.091(\pm 0.433)~,
\end{equation} 
and the red peaks:\begin{equation}
    (g'-i')_0=0.011(\pm 0.019)*g'_0+0.746(\pm 0.458)~.
\end{equation}

We note that in the brightest magnitude bin, although the Akaike Information Criterion points toward a unimodal color distribution (Tab.~\ref{tab:GMM}), the two calculated peaks for the bimodal case seems to fall along the linear fits for the two populations (Fig.~\ref{fig:cmd 3894}). In order to test our results, we then repeated the linear fits with the addition of these two points. We note that the best fit parameters are not significantly different from the previous ones, but the addition of these two points result in a decrease in the errors of the parameters. In particular, for the blue population we have $m=-0.011\pm 0.009$ and $q=1.022\pm 0.215$, while for the red population $m=0.013\pm 0.009$ and $q=0.691\pm 0.227$, where $m$ and $q$ are the slope and intercept of the fitted line.

\begin{table*}
\caption{Best-fit parameters for the Gaussian mixture modeling of the $(g'-i')_0$ color distribution of GC candidates.}
    \centering
    \begin{tabular}{c c c c c c c c}
    \hline
    $g'_{\rm 0}$ & $N_{\rm GC}$ &  $\mu_{\rm b}$ & $\mu_{\rm r}$ & $\sigma_{\rm b}$ & $\sigma_{\rm r}$ & Bootstrap(DD) & $N$ \\
    \hline     
    $22.0 - 26.789$ & $3818$ & $0.763\pm 0.004$ & $1.012\pm 0.004$ & $0.151\pm 0.004$ & $0.107\pm 0.008$ & $2.15\pm 0.12$ & $2$\\
    $22.0 - 23.47$ & $399$ & $0.765\pm 0.008$ & $0.981\pm 0.010$ & $0.119\pm 0.010$ & $0.130\pm 0.010$ & $2.39\pm 0.58$ & $1$\\
    $23.47 - 24.023$ & $400$ & $0.753\pm 0.007$ & $1.004\pm 0.009$ & $0.110\pm 0.009$ & $0.123\pm 0.009$ & $2.49\pm 0.26$ & $2$\\
    $24.023 - 24.415$ & $401$ & $0.756\pm 0.010$ & $0.987\pm 0.09$ & $0.119\pm 0.014$ & $0.120\pm 0.013$ & $2.00\pm 0.71$ & $2$ \\
    $24.415 - 24.795$ & $400$ & $0.765\pm 0.009$ & $1.028\pm 0.008$ & $0.115\pm 0.009$ & $0.120\pm 0.009$ & $2.59\pm 0.25$ & $2$ \\
    $24.795 - 25.085$ & $401$ & $0.714\pm 0.009$ & $0.987\pm 0.009$ & $0.118\pm 0.010$ & $0.135\pm 0.010$ & $2.30\pm 0.32$ & $2$ \\
    $25.085 - 25.350$ & $400$ & $0.748\pm 0.008$ & $1.025\pm 0.010$ & $0.128\pm 0.008$ & $0.128\pm 0.008$ & $2.18\pm 0.24$ & $2$ \\
    $25.350 - 25.635$ & $400$ & $0.770\pm 0.010$ & $1.008\pm 0.009$ & $0.128\pm 0.010$ & $0.140\pm 0.009$ & $2.35\pm 0.73$ & $1$ \\
    $25.635 - 25.944$ & $401$ & $0.779\pm 0.010$ & $1.033\pm 0.009$ & $0.139\pm 0.008$ & $0.140\pm 0.008$ & $2.34\pm 0.35$ & $1$ \\
    $25.944 - 26.376$ & $400$ & $0.820\pm 0.010$ & $1.077\pm 0.009$ & $0.140\pm 0.011$ & $0.130\pm 0.011$ & $2.17\pm 0.64$ & $3$ \\
    $26.376 - 26.789$ & $216$ & $0.758\pm 0.015$ & $1.013\pm 0.014$ & $0.159\pm 0.013$ & $0.144\pm 0.014$ & $2.99\pm 0.50$ & $1$\\
    \hline
    \end{tabular}
    \tablefoot{The columns show the magnitude bin ($g'_0$), the number of sources in the bin ($N_{\rm GC}$), the positions of the blue and red peaks ($\mu_{\rm b}$ and $\mu_{\rm r}$), the sigma of the two Gaussians ($\sigma_{\rm b}$ and $\sigma_{\rm r}$), the separation of the peaks (Bootstrap(DD), defined as the separation of the means relative to their widths), and the number of components defined by the Akaike Information Criterion ($N$).}
    \label{tab:GMM}
\end{table*}

\subsection{Radial trends of the color distribution}
\begin{figure}
    \centering
    \includegraphics[scale=0.25]{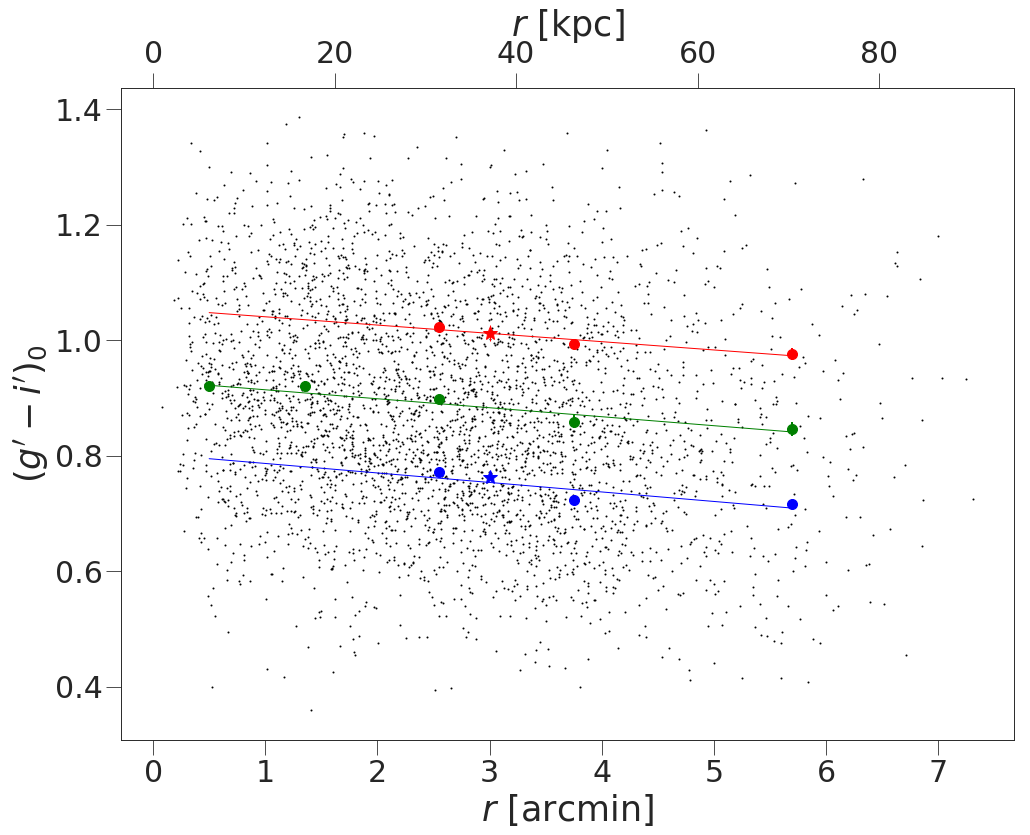}
    \caption{$(g'-i')_0$ color versus radius plot. The black dots represent the $3818$ GC candidates. The blue and red dots are the blue and red peaks of the color distributions divided into bins as explained in this Section. The green dots represent the medians between the blue and red peaks. The blue, red, and green lines are the linear fit of the peaks. The blue and red stars represent the peaks of the total color distribution. The effective radius of the galaxy is $r_{\rm eff}=1.7'$, which at the distance of NGC~4696 corresponds to $21.02$~kpc.}
    \label{fig:linfit_rad}
\end{figure}

\begin{figure*}
    \centering
    \includegraphics[scale=0.27]{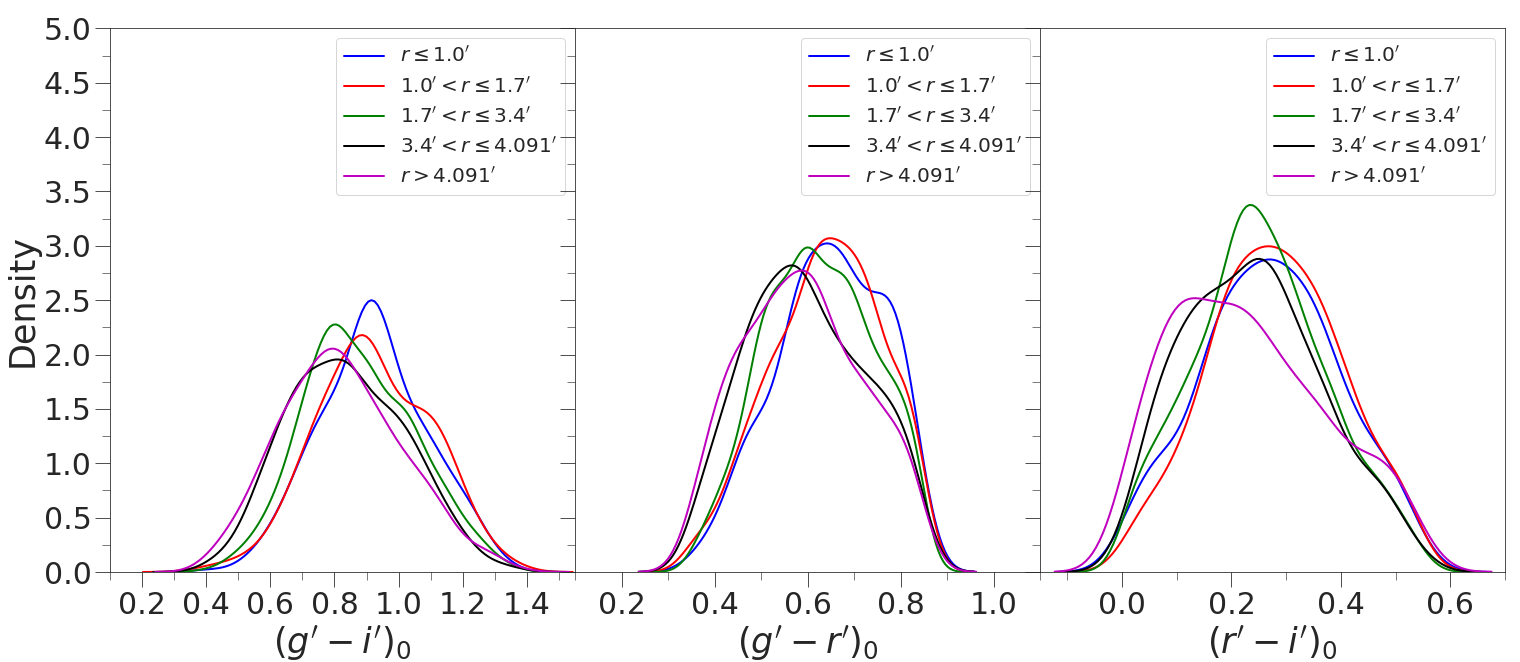}
    \caption{$(g'-i')_0$, $(g'-r')_0$, and $(r'-i')_0$ color distributions for the globular cluster subsamples in different radial bins. The number of sources in the different bins is $N_{\rm GC}=428$ (blue), $N_{\rm GC}=634$ (red), $N_{\rm GC}=1638$ (green), $N_{\rm GC}=540$ (black), and $N{\rm GC}=578$ (magenta).}
    \label{fig:reff bin}
\end{figure*}

In order to study eventual color gradients across the galaxy, we divided the GC sample in subsamples of radial bins, and calculated the color distribution for the $(g'-i')_0$, $(r'-i')_0$, and $(g'-r')_0$ colors in each bin. In particular, at $r\leq 1'$ a visual inspection shows what appears to be a sample of clusters more concentrated toward a single, intermediate value of the $(g'-i')_0$ color (Fig.~\ref{fig:linfit_rad}). The Akaike Information Criterion shows (Tab.~\ref{tab:rad_bin}) that for the innermost bin and the one with $1'<r\leq 1.7'$, the subsamples are better described by unimodal color distributions, with peaks of the Gaussians in both cases being at $(g'-i')_0=0.921$. However, data for $r\leq 1$~arcmin should be taken with caution as the high noise associated with the galaxy body hampers a reliable detection and characterization of GCs in this region. In the three outermost bins, the color distributions appear bimodal, with both the blue and red peaks shifting toward bluer colors as we go toward larger galactocentric distances. The results are shown in Fig.~\ref{fig:reff bin} and in Tab.~\ref{tab:rad_bin}. From this analysis, the color distribution shows a trend with radius, such as it has been observed for other early type galaxies (\citealt{Bassino}). To better quantify this trend, we performed a linear fit of the blue and red peaks, as well as on the medians between the two peaks which represent the color separation between the two populations. The linear fits are shown in Fig.~\ref{fig:linfit_rad} and the best fit parameters are reported in Tab.~\ref{tab:rad_bin_linfit}. 

Assuming that our GC candidates are all coeval and old, to check our results, we converted the colors in metallicities, such as (\citealt{Faifer11}):\begin{equation}
    [\rm{Z/H}]=3.51*(g'-i')_0-3.91~.
\end{equation}
The chemical abundances $[\rm{Z/H}]$ is related to the iron abundance by (\citealt{Mendel07}):\begin{equation}
    [\rm{Fe/H}]=[\rm{Z/H}]-0.131~.
\end{equation}
We calculated the values of $[\rm{Fe/H}]$ for the median between the blue and red peaks ($\mu_{\rm med}$ in Tab.~\ref{tab:rad_bin}), and fitted them with \begin{equation}
    [\rm{Fe/H}]=m*\log\left(\frac{r}{r_{\rm eff}}\right)+q~,
\end{equation}
where $r$ is the radius and $r_{\rm eff}$ is the effective radius. Our best fit parameters, $m=-0.265\pm 0.077$ and $q=-0.899\pm 0.029$, are in good agreement with those obtained in the photometric survey of GCSs in brightest cluster ellipticals by \citet{Harris23}. We are aware that some theoretical models show the possibility that the color-metallicity relation could be nonlinear (e.g., \citealt{Yoon06}), with a bimodal color distribution resulting from a unimodal metallicity distribution. However, the color-metallicity relations are only modestly nonlinear for optical color indices. Moreover, several spectroscopic studies found a bi- or even trimodal metallicity distribution for the GCSs analyzed (e.g., \citealt{Caldwell16}; \citealt{Villaume19}). Finally, despite the test of different types of relations, such as linear, two parts piecewise linear correlations, and continuous quadratic relations, no consensus exists yet on the color-metallicity relation (\citealt{Harris23}). Our transformations are based on \citet{Faifer11} as described above. The positions of the peaks of the metallicity distribution inferred from these transformations are at $[\rm{Fe/H}]=-1.363\pm 0.010$ and $[\rm{Fe/H}]=-0.488\pm 0.012$, in good agreement with the values found by \citealt{Harris23}.

\begin{table*}
\caption{Best-fit parameters for the Gaussian mixture modeling of the $(g'-i')_0$ color distribution of GC candidates divided into radial bins.}
    \centering
    \begin{adjustbox}{width=\textwidth}
    \begin{tabular}{c c c c c c c c c}
    \hline
     $r$ & $N_{\rm GC}$ & $\mu_{\rm b}$ & $\mu_{\rm r}$ & $\sigma_{\rm b}$ & $\sigma_{\rm r}$ & $\mu_{\rm med}$ & Boot(DD) & $f_{\rm blue}$ \\
    \hline
    $\leq 1'$ & $428$ & $-$ & $-$ & $-$ & $-$ & $0.921\pm 0.008$ & $2.17\pm 1.08$ & $-$ \\
    $1'<r\leq 1.7'$ & $634$ & $-$ & $-$ & $-$ & $-$ & $0.921\pm 0.009$ & $2.11\pm 0.58$ & $-$ \\
    $1.7'<r\leq 3.4'$ & $1638$ & $0.772\pm 0.006$ & $1.023\pm 0.004$ & $0.117\pm 0.007$ & $0.130\pm 0.007$ & $0.898\pm 0.006$ & $2.19\pm 0.18$ & $1.42$ \\
    $3.4'<r\leq 4.091'$ & $540$ & $0.724\pm 0.008$ & $0.994\pm 0.007$ & $0.121\pm 0.008$ & $0.126\pm 0.008$ & $0.859\pm 0.011$ & $2.04\pm 0.26$ & $1.88$\\
    $>4.091'$ & $578$ & $0.716\pm 0.007$ & $0.976\pm 0.010$ & $0.130\pm 0.011$ & $0.148\pm 0.011$ & $0.846\pm 0.012$ & $2.04\pm 0.43$ & $2.14$ \\
    \hline
    \end{tabular}
    \end{adjustbox}
    \tablefoot{The columns show the radial bin, the number of GCs in the bin ($N_{\rm GC}$), the positions of the blue and red peaks ($\mu_{\rm b}$ and $\mu_{\rm r}$), the sigma of the two Gaussians ($\sigma_{\rm b}$ and $\sigma_{\rm r}$), the position of the separation between the blue and red populations ($\mu_{\rm med}$), the Bootstrap(DD) parameter (Boot(DD)), and the fraction of blue clusters ($f_{\rm blue}$).}
    \label{tab:rad_bin}
\end{table*}

\begin{table}
\caption{Best-fit parameters for the linear interpolation of the blue, red, and median peaks of the $(g'-i')_0$ distribution divided into radial bins as in Tab.~\ref{tab:rad_bin}.}
    \centering
    \begin{tabular}{c c c}
    \hline
      & $m$ & $q$  \\
    \hline
    Red & $-0.014\pm 0.004$ & $1.055\pm 0.017$ \\
    Blue & $-0.016\pm 0.010$ & $0.803\pm 0.040$ \\
    Median & $-0.016\pm 0.007$ & $0.930\pm 0.029$ \\
    \hline
    \end{tabular}
    \tablefoot{The lines show the values for the red, blue, and median (green) populations. $m$ and $q$ are the slope and intercept of the fitted lines, respectively.}
    \label{tab:rad_bin_linfit}
\end{table}

\subsection{Spatial distribution}
\label{sec:spatial distr}
\begin{figure}
    \centering
    \includegraphics[scale=0.18]{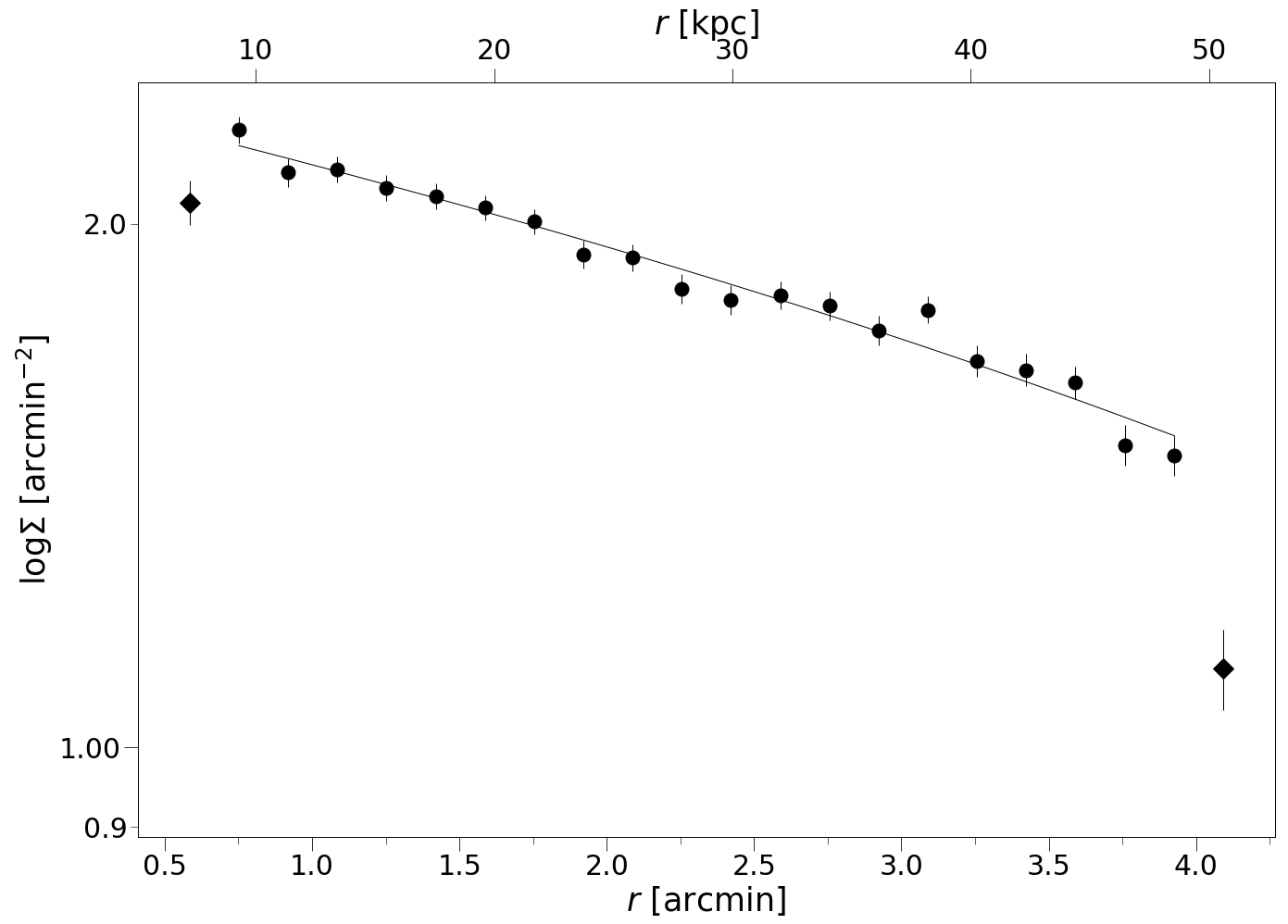}
    \caption{Radial profile of the clusters' surface density. The plot shows the density of GCs ($\log \Sigma$) as a function of the distance to the center of the galaxy ($r$). The diamond symbols represent data not included in the regression (indicated by the solid line), in order to minimize incompleteness effects.}
    \label{fig:density profile}
\end{figure}

We calculated the radial profile of the clusters' number surface density by counting the number of GC candidates inside circular annuli centered at the optical center of the galaxy. We chose a width of $0.167$~arcmin for each annulus. Then, we calculated the density of GCs in each bin. In order to account for the spatial limits of the image, we rerun {\tt ISOFIT} holding the center fixed, using a minimum semi-major axis length of $20.04$~arcsec (since in our sample there are no sources inside this radius), and adopting the same bin size as indicated above. The parameter {\tt NPIX$\_$C} in the output table gives the total number of valid pixels inside a circle having radius equal to the semi-major axis calculated by the interpolation (\citealt{jed}). Then the areas of the circular annuli were computed and used to estimate the average surface density at that mean radius. The calculated areas are already corrected for any geometrical incompleteness due to the position of the galaxy in the image since {\tt NPIX$\_$C} gives only the available pixels. Since the innermost bins (up to $r=0.585$~arcmin) are affected by incompleteness effects due to the difficulty of identifying sources in the brightest parts of the galaxy, we decided to exclude them from the analysis. The upper limit in radius, see Fig.~\ref{fig:positions}, was given by the field of view of our central pointing. Our sample is then limited to the clusters located at a distance from the center of the galaxy of $0.585<r<4.091$~arcmin ($0.344~r_{\rm eff}<r<2.406~r_{\rm eff}$). In this radial range, the number of GC candidates is $N_{\rm GC}=3083$. The result is shown in Fig.~\ref{fig:density profile}, where the diamond symbols indicate the innermost and outermost bins excluded from the analysis, where the sharp drop in the outermost bin is given by geometric incompleteness. The remaining distribution follows a linear relation of the type \textbf{\begin{equation}
\log\Sigma=m*r+q
\end{equation}}
in the $\log\Sigma-r$ plane. The parameters of the fit are reported in Tab. \ref{tab:linear fit}. The calculated areas and GC densities are reported in Tab.~\ref{tab:area}.

\begin{table}
\caption{Best-fit parameters for the linear interpolation of the radial density profile of GC candidates in the $\log\Sigma-r$ plane.}
    \centering
    \begin{tabular}{c c c}
    \hline
         &  $m$ & $q$\\
    \hline     
    Total & $-0.219\pm 0.009$ & $2.379\pm 0.022$ \\
    Red & $-0.279\pm 0.012$ & $2.145\pm 0.030$ \\
    Blue & $-0.174\pm 0.012$ & $2.029\pm 0.029$ \\
    \hline
    \end{tabular}
    \tablefoot{The rows show the values for the total, red, and blue samples. $m$ is in arcmin.}
    \label{tab:linear fit}
\end{table}

\begin{table}
\caption{Radial number density profile for the GC candidates.}
    \centering
    \begin{tabular}{c c c c}
    \hline
    $r$ & $A$ & $N_{\rm GC}$ & $\Sigma$ \\
    (arcmin) & (arcmin$^2$) & & (arcmin$^{-2}$) \\
    \hline
  0.585 & 0.438 & 50 & 114$\pm$16\\
  0.752 & 0.614 & 113 & 184$\pm$17\\
  0.919 & 0.788 & 109 & 138$\pm$13\\
  1.086 & 0.964 & 136 & 141$\pm$12\\
  1.252 & 1.139 & 143 & 125$\pm$10\\
  1.420 & 1.314 & 156 & 119$\pm$10\\
  1.587 & 1.490 & 165 & 111$\pm$9\\
  1.754 & 1.664 & 169 & 102$\pm$8\\
  1.921 & 1.839 & 153 & 83$\pm$7\\
  2.088 & 2.015 & 165 & 82$\pm$6\\
  2.255 & 2.191 & 150 & 68$\pm$6\\
  2.421 & 2.365 & 152 & 64$\pm$5\\
  2.589 & 2.541 & 168 & 66$\pm$5\\
  2.756 & 2.715 & 169 & 62$\pm$5\\
  2.923 & 2.892 & 158 & 55$\pm$4\\
  3.090 & 3.067 & 187 & 61$\pm$4\\
  3.257 & 3.241 & 151 & 47$\pm$4\\
  3.424 & 3.418 & 152 & 44$\pm$4\\
  3.591 & 3.592 & 150 & 42$\pm$3\\
  3.758 & 3.768 & 117 & 31$\pm$3\\
  3.925 & 3.942 & 117 & 30$\pm$3\\
  4.091 & 4.117 & 53 & 13$\pm$2\\
  \hline
    \end{tabular}
    \tablefoot{The columns show the radius of the central annulus point ($r$), the area of the circular annulus ($A$), the number of clusters inside the circular annulus ($N_{\rm GC}$), and the density of clusters (\textbf{$\Sigma=N_{\rm GC}/A$}).}
    \label{tab:area}
\end{table}

\subsection{Blue versus red population}
\label{blue vs red}

\begin{table}
\caption{Radial number density profiles for the red and blue GC populations.}
    \centering
    \resizebox{\columnwidth}{!}{\begin{tabular}{c c c c c c c}
    \hline
     $r$ & $A$ & $N_{\rm red}$ & $N_{\rm blue}$ & $\Sigma_{\rm red}$ & $\Sigma_{\rm blue}$ & $N_{\rm blue}/N_{\rm red}$ \\
     (arcmin) & (arcmin$^2$) & & & (arcmin$^{-2}$) & (arcmin$^{-2}$) & \\
     \hline
    0.585 & 0.438 & 26 & 24 & 59$\pm$11 & 55$\pm$11 & 1.083\\
  0.752 & 0.614 & 51 & 62 & 83$\pm$12 & 101$\pm$13 & 1.26\\
  0.919 & 0.788 & 54 & 55 & 68$\pm$9 & 70$\pm$9 & 1.094\\
  1.086 & 0.964 & 68 & 68 & 71$\pm$9 & 71$\pm$8 & 1.015\\
  1.252 & 1.139 & 74 & 69 & 64$\pm$8 & 61$\pm$7 & 0.973\\
  1.420 & 1.314 & 85 & 71 & 65$\pm$7 & 54$\pm$6 & 0.915\\
  1.587 & 1.490 & 77 & 88 & 52$\pm$6 & 59$\pm$6 & 1.253\\
  1.754 & 1.664 & 87 & 82 & 52$\pm$6 & 49$\pm$5 & 0.955\\
  1.921 & 1.839 & 70 & 83 & 38$\pm$5 & 45$\pm$5 & 1.328\\
  2.088 & 2.015 & 75 & 90 & 37$\pm$4 & 45$\pm$5 & 1.267\\
  2.255 & 2.191 & 63 & 87 & 29$\pm$4 & 40$\pm$4 & 1.435\\
  2.421 & 2.365 & 57 & 95 & 24$\pm$3 & 40$\pm$4 & 1.786\\
  2.589 & 2.541 & 67 & 101 & 26$\pm$3 & 40$\pm$4 & 1.522\\
  2.756 & 2.715 & 60 & 109 & 22$\pm$3 & 40$\pm$4 & 1.915\\
  2.923 & 2.892 & 66 & 92 & 23$\pm$3 & 32$\pm$3 & 1.403\\
  3.090 & 3.067 & 73 & 114 & 24$\pm$3 & 37$\pm$3 & 1.595\\
  3.257 & 3.241 & 61 & 90 & 19$\pm$2 & 28$\pm$3 & 1.550\\
  3.424 & 3.418 & 58 & 94 & 17$\pm$2 & 28$\pm$3 & 1.672\\
  3.591 & 3.592 & 46 & 104 & 13$\pm$2 & 29$\pm$3 & 2.188\\
  3.758 & 3.768 & 41 & 76 & 11$\pm$2 & 20$\pm$2 & 1.860\\
  3.925 & 3.942 & 42 & 75 & 11$\pm$2 & 19$\pm$2 & 1.733\\
  4.091 & 4.117 & 18 & 35 & 4$\pm$1 & 9$\pm$1 & 2.118\\
    \hline
    \end{tabular}}
    \tablefoot{The columns show the radius of the central annulus point ($r$) calculated as the average radius between two consecutive annuli, the area of the circular annulus ($A$), the number of red clusters ($N_{\rm red}$), the number of blue clusters ($N_{\rm blue}$), the density of red clusters (\textbf{$\Sigma_{\rm red}=N_{\rm red}/A$}), the density of blue clusters (\textbf{$\Sigma_{\rm blue}=N_{\rm blue}/A$}), and the ratio of blue to red clusters ($N_{\rm blue}/N_{\rm red}$).}
    \label{tab:blue to red}
\end{table}

\begin{figure}
    \centering
    \includegraphics[scale=0.18]{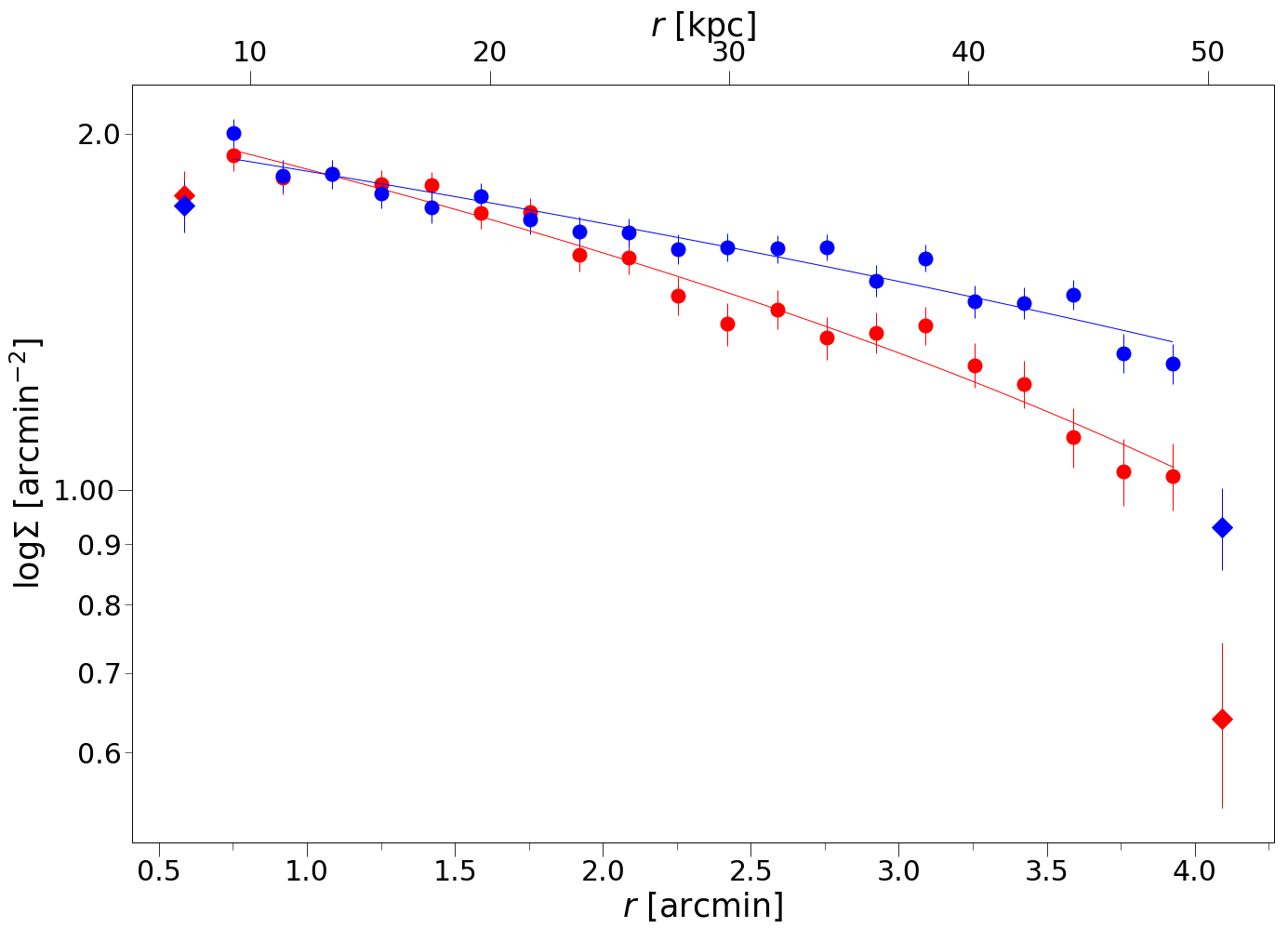}
    \caption{Radial density profile for the blue and red GC populations. The plot shows the logarithm of the density of clusters ($\log \Sigma$) as a function of the distance from the center of the galaxy ($r$) for the $(g'-i')_0\geq 0.905$ (red) and $(g'-i')_0<0.905$ (blue) subsamples. The diamonds represent the data not considered for the analysis in order to avoid incompleteness effects.}
    \label{fig:blue red linear fit}
\end{figure}

We divided the sample into the red and blue GC populations, and calculated the radial density profiles. For the division between the blue and red clusters we adopted the fixed value of $(g'-i')_0=0.905$~mag (see Sec.~\ref{sec:color distr}). In this way, we have $N_{\rm red}=1319$ and $N_{\rm blue}=1764$. The results are shown in Fig.~\ref{fig:blue red linear fit}, where the diamond symbols indicate the innermost and outermost bins excluded from the analysis. As can be seen, the two populations are well described by linear relations, with different slopes. The parameters of the linear fits are reported in Tab.~\ref{tab:linear fit}.

\begin{figure}
    \centering
    \includegraphics[scale=0.18]{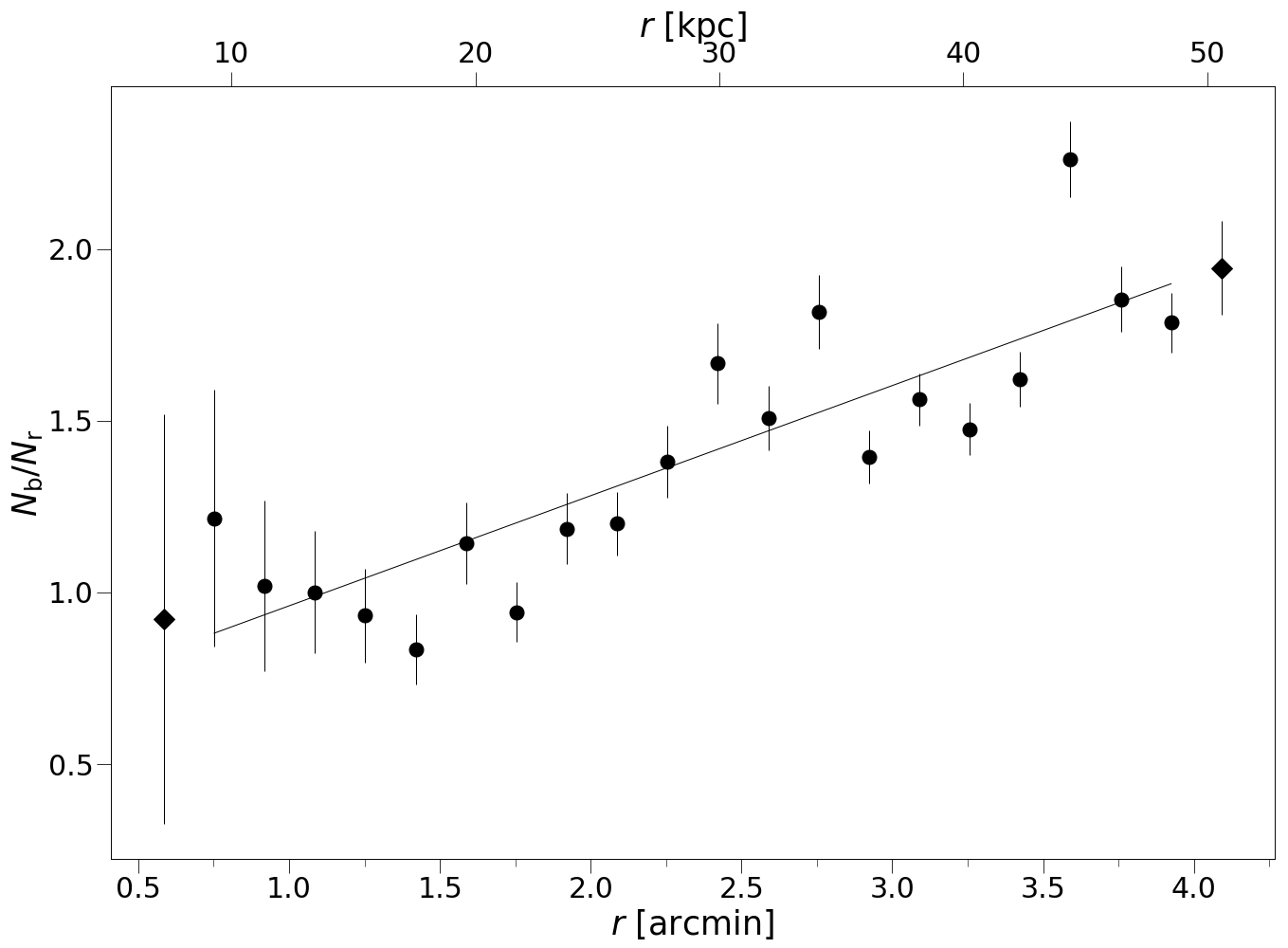}
    \caption{Ratio between the blue and red populations of globular clusters. We considered radial bins of $0.167$ arcmin, as in the radial density plots. The plot shows the ratio of blue to red clusters ($N_{\rm blue}/N_{\rm red}$) as a function of the distance from the center of the galaxy ($r$). The diamonds represent data not included in the study in order to avoid incompleteness effects in the innermost bin. The line shows the result of the linear fit to the data.}
    \label{fig:ratio_br}
\end{figure}

We finally calculated the ratio between the blue and red clusters, keeping the same radial bin used for the radial density profiles. The calculated densities and ratios for each bins are reported in Tab.~\ref{tab:blue to red} and the result is shown in Fig.~\ref{fig:ratio_br}. As can be seen, the ratio of the blue to red population tends to increase going to larger galactocentric distances. This was expected since the red GCs are usually more centrally concentrated with respect to the blue ones, and it is clearly visible in Fig.~\ref{fig:blue red linear fit}.

\subsection{The azimuthal distribution of the globular cluster candidates}

We studied the angular distribution of the GC candidates in order to determine if they are distributed with some preferential direction, since this is an additional test to probe the interaction and merger history of NGC~4696. In particular, the comparison with the positions of the galaxies NGC~4709 and NGC~4696B (see Tab.~\ref{tab:galaxies parameters}) is important since \citet{Walker} outlined that the X-ray emission shows a temperature excess in the direction of NGC~4709 and a filamentary structure extending to NGC~4696B. The angular distances and directions of the two galaxies with respect to NGC~4696, calculated using Aladin\footnote{Available in https://aladin.u-strasbg.fr/aladin.gml} (\citealt{aladin1}; \citealt{aladin2}) by drawing a distance vector between the centers of the galaxies, are shown in Tab.\ref{tab:galaxies parameters}. We tested the results by calculating the PA using the coordinates package in Astropy (\citealt{astropy1}; \citealt{astropy2}; \citealt{astropy3}), which given the coordinates of two objects calculates the position angle of one with respect to the other. The calculation gave similar results to those obtained with Aladin ($107.1^{\circ}$ for NGC~4709 and $284.9^{\circ}$) for NGC~4696B.
 
\begin{figure}
    \centering
    \includegraphics[scale=0.26]{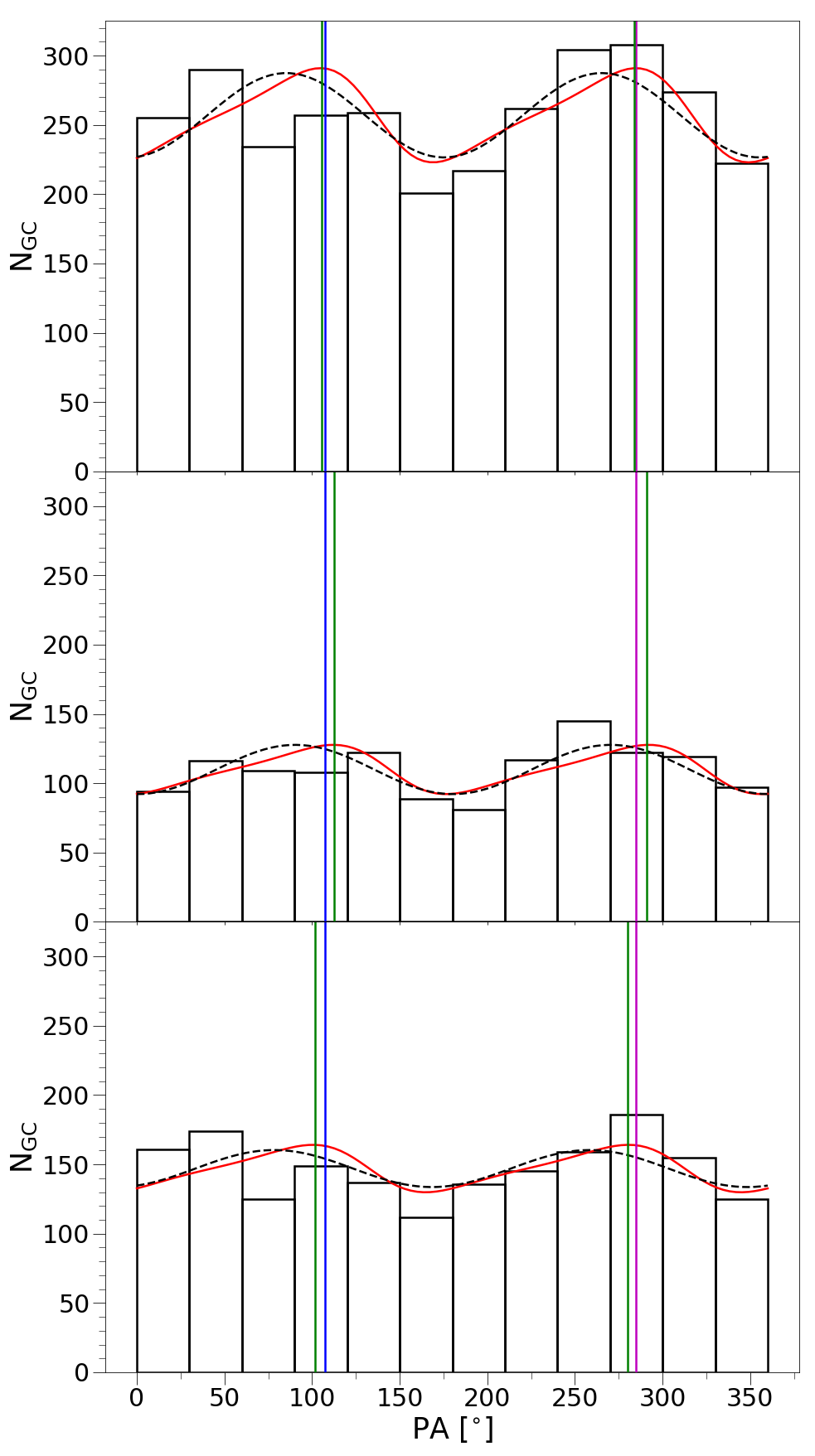}
    \caption{Azimuthal distribution of $3083$ globular cluster candidates. The plots show the number of clusters ($N_{\rm GC}$) as a function of the position angle (PA), divided in angular bins of $30^{\circ}$. The red line represents the interpolation with the function described in Eq.~\ref{eq:asym} to the data, the blue and magenta lines represent the angular positions of NGC~4709 and NGC~4696B respectively, the green lines represent the position angle of the peaks, and the dashed black lines represent the fit with Eq.~\ref{eq:sin}. The different panels show the distributions of the total ({\em top panel}), red ({\em middle panel}), and blue ({\em lower panel}) GC samples.}
    \label{fig:total_pa}
\end{figure}

 As for the radial number density profile, we limited our study to the annulus defined by $0.585\leq r\leq 4.091$~arcmin. With this radial selection we obtained a symmetrical region around the galaxy. The angular distribution of the total, red and blue distributions is shown in Fig.\ref{fig:total_pa}, where all the samples are divided in angular bins of $\theta=30^{\circ}$. The angular positions of the GCs are measured eastward starting from the north axis. 

\begin{table}
\caption{Parameters of the galaxies NGC~4696B and NGC~4709.}
\large
    \centering
    \resizebox{\columnwidth}{!}{\begin{tabular}{c c c c c c}
    \hline
    Name & RA & DEC & $d$ & PA & $e$ \\
     & (J2000) & (J2000) & (arcmin) & ($^{\circ}$) & \\
     \hline
     NGC~4696B & $12^{\rm h}47^{\rm m}21^{\rm s}.78$ & $-41^{\circ}14'15".0$ & $17.06$ & $284.8$ & $0.39$ \\
     NGC~4709 & $12^{\rm h}50^{\rm m}03^{\rm s}.88$ & $-41^{\circ}22'55".10$ & $14.5$ & $107.2$ & $0.167$ \\
     \hline
    \end{tabular}}
    \tablefoot{The columns report the name of the object, the coordinates tabulated in NED, the distance ($d$) from NGC~4696, the angular position with respect to the center of NGC~4696 calculated eastward starting from the North direction (PA), and the ellipticity tabulated in NED.}
    \label{tab:galaxies parameters}
\end{table}

As can be seen, the distributions appear to have a sinusoidal shape, so we used optimize package of SciPy (\citealt{scipy}) to determine the best fit to the data. In particular, we noticed that the azimuthal distribution looks more complex than a simple sinusoidal. In order to take into account for the asymmetric shape of the distribution, we fit it with the function \begin{equation}
    y=a+b*\sin\left[\frac{m*\pi *(x-c)}{180}-0.5*\sin\left(\frac{m*\pi(x-c)}{180}\right)\right]~,
    \label{eq:asym}
\end{equation}
and we compared the results with a symmetrical sinusoidal function given by\begin{equation}
    y=a+b*\sin\left[\frac{2\pi *(x-c)}{180}\right]~,
    \label{eq:sin}
\end{equation}
where $y$ is the number of clusters, $x$ is the position angle, $a$ is the vertical shift of the function, $b$ is the amplitude, and $c$ is the phase shift. After some careful attempts of fitting, we decided to fix the value of $m$. The best fit parameters and the positions of the two peaks of the sinusoidal fit for the total, red and blue distributions are reported in Tab.~\ref{tab:best fit}. It is important to note that the peaks found with the fit of the asymmetrical sinusoidal are much closer to the directions of NGC~4709 and NGC~4696B with respect to those obtained with the simple sinusoidal fit. 

The sinusoidal shape of the GC counts is typical of an elliptical distribution of the GCS. In order to calculate the ellipticity, we performed the 2D Kernel Density Estimation of {\tt sklearn.neighbors} (\citealt{Ped}) to create a grid of the sample locations using the physical positions of the GC candidates in the image, and used the results to fit a Bivariate Gaussian Distribution using a routine in astroML (\citealt{astroML}). The outputs of this routine are the position of the center of the ellipse, the sigmas of the two Gaussians which can be used as major and minor axis to calculate the ellipticity, and the position angle. The results of the routine within a $68\%$ confidence interval are shown in Tab.~\ref{tab:shape}.

\begin{table}
\caption{Azimuthal distribution of the GC candidates.}
    \centering
    \resizebox{\columnwidth}{!}{\renewcommand{\arraystretch}{1.5}\begin{tabular}{c c c c c c c c}
    \hline
         & $a$ & $b$ & $c$ & $m$ & $y_{\rm peak}$ & PA$_1 [^{\circ}]$ & PA$_2 [^{\circ}]$ \\
    \hline    
     Total$_{\rm Eq.\ref{eq:asym}}$ & $256.93^{+1.09}_{-1.05}$ & $33.96^{+1.63}_{-1.62}$ & $52.13^{+4.29}_{-6.09}$ & $2$ & $290.87^{+2.33}_{-2.40}$ & $105.17\pm 0.28$ & $284.34^{+2.93}_{-0.70}$ \\
     Red$_{\rm Eq.\ref{eq:asym}}$ & $109.98^{+0.52}_{-0.50}$ & $17.81\pm 0.74$ & $54.12^{+0.90}_{-0.91}$ & $2$ & $127.77^{+1.07}_{-1.06}$ & $112.30\pm 0.43$ & $291.67^{+2.88}_{-0.76}$ \\
     Blue$_{\rm Eq.\ref{eq:asym}}$ & $147.00^{+0.72}_{-0.69}$ & $17.06^{+1.21}_{-1.18}$ & $42.61^{+1.01}_{-0.96}$ & $2$ & $164.04^{+1.60}_{-1.64}$ & $100.71^{+1.11}_{-2.52}$ & $280.28\pm 0.28$ \\
     Total$_{\rm Eq.\ref{eq:sin}}$ &  $256.94^{1.10}_{-1.12}$ & $30.44^{+1.71}_{-1.72}$ & $40.22^{+1.19}_{-1.18}$ & $-$ & $287.37^{+2.48}_{-2.47}$ & $85.22^{+1.64}_{-1.39}$ & $265.22^{+1.64}_{-1.39}$ \\
     Red$_{\rm Eq.\ref{eq:sin}}$ & $109.93^{+0.51}_{-0.50}$ & $17.78^{+0.74}_{-0.72}$ & $45.39^{+1.05}_{-1.07}$ & $-$ & $127.71^{+1.14}_{-1.10}$ & $90.38^{+1.54}_{-1.49}$ & $270.38^{+1.54}_{-1.49}$ \\
     Blue$_{\rm Eq.\ref{eq:sin}}$ & $147.00^{+0.82}_{-0.83}$ & $13.42^{+1.26}_{-1.29}$ & $33.33^{+2.11}_{-2.59}$ & $-$ & $160.42^{+1.73}_{-1.69}$ & $78.34^{+2.46}_{-2.59}$ & $258.34^{+2.46}_{-2.59}$ \\
     \hline
    \end{tabular}}
    \tablefoot{Best-fit parameters and peak values of the sinusoidal interpolation performed with SciPy for the total, red and blue GC samples, using Eq.\ref{eq:asym} and Eq.\ref{eq:sin}. The errors were calculated via the Bootstrap method.}
    \label{tab:best fit}
\end{table}

\begin{table}
\caption{Shape of the GCS.}
\fontsize{20pt}{20pt}\selectfont
    \centering
    \resizebox{\columnwidth}{!}{\renewcommand{\arraystretch}{1.5}\begin{tabular}{c c c c c c c}
    \hline
         & RA & DEC & $a$ & $b$ & PA & $e$ \\
         & (J2000) & (J2000) & (arcmin) & (arcmin) & $[^{\circ}]$ & \\
    \hline     
     Total &  $12^{\rm h}48^{\rm m}49^{s}.25^{+0.40}_{-0.39}$ & $-41^{\circ}18'37''.27^{+3.88}_{-3.47}$ & $3.709^{+0.068}_{-0.076}$ & $2.901^{+0.042}_{-0.049}$ & $90$ & $0.218^{+0.020}_{-0.021}$ \\
     Red & $12^{\rm h}48^{\rm m}49^{\rm s}.25^{+0.74}_{-0.66}$ & $-41^{\circ}18'30''.85^{+14.02}_{-13.36}$ & $3.446^{+0.134}_{-0.124}$ & $2.638^{+0.178}_{-0.169}$ & $90$ & $0.233^{+0.059}_{-0.058}$ \\
     Blue & $12^{\rm h}48^{\rm m}49^{\rm s}.07^{+0.070}_{-0.065}$ & $-41^{\circ}18'37''.41^{+10.67}_{-10.04}$ & $3.546^{+0.129}_{-0.123}$ & $2.720\pm 0.124$ & $90$ & $0.232^{+0.044}_{-0.045}$ \\
     \hline
    \end{tabular}}
    \tablefoot{Parameters of the elliptical shape of the GCS for the total, red, and blue populations. The columns report the right ascension and declination of the center of the ellipse (RA and DEC), the major and minor axis ($a$ and $b$), the position angle (PA), and the ellipticity ($e$). The errors on the parameters are calculated via the Bootstrap method.}
    \label{tab:shape}
\end{table}

The results of the fit are shown in Fig.~\ref{fig:total_pa} for the total, red and blue distributions.  As can be seen, all distributions appear to be sinusoidal, with the peaks positions (green dashed lines) equal or extremely close to the directions of the galaxies NGC~4709 and NGC~4696B, represented by the blue and magenta lines. 

\subsection{Comparison with NGC~4696 profile}

In order to have a preliminary idea of the radial surface brightness profile of NGC~4696, we rerun {\tt ISOFIT} with the same parameters as before but keeping the position of the galaxy's center fixed. This allows to highlight eventual asymmetries in the isophotes' shapes. Through the Fourier coefficients $a_3$, $a_4$, $b_3$, and $b_4$, we can quantify the deviations from the elliptical shape of the isophotes (\citealt{ciambur}). The $b_4$ coefficient results particularly important: positive values describe a disky isophote, whereas negative values describe a boxy isophote. The surface brightness profiles along the major axis, the position angles, ellipticities, and the values of the Fourier coefficients are shown in Fig.~\ref{fig:ngc profile}. Our profiles are spatially limited to a galactocentric radius of $\sim 2.4$~arcmin, so they do not reach the outermost regions of the halo. From these preliminary results, we can see some strong fluctuations in the values of the position angle and ellipticity of NGC~4696, from the inner to the outer regions. This could be due to the presence of the dust lanes in the central regions of the galaxy, and to past interactions with other galaxies in the cluster. Moreover, we see that there is a change in the value of the parameter $b_4$, which goes from negative values in the inner regions to positive values in the outer regions. This suggests that some outer isophotes have a disky shape, a feature that will be further analyze in a future paper (Federle et al., in prep.). Comparing the values with those found for the shape of the GCS (Tab.~\ref{tab:shape}), we can see that NGC~4696 appear to have a lower ellipticity and position angle with respect to the total, red and blue GCSs. In particular, we calculated the mean value for the ellipticity and its standard deviation in a sample of $10$ points around the half-light radius of NGC~4696, which, from our spatially limited analysis, is inferred to be located at a radius of $r\sim 0.45$~arcmin. The ellipticity of the galaxy results of $e=0.135\pm 0.007$, which is lower than the values of $e=0.218^{+0.020}_{-0.021}$, $e=0.233^{+0.059}_{-0.058}$, and $0.232^{+0.044}_{-0.045}$ and found for the total, red and blue GCSs. These differences point toward a disturbed GCS, with a shape that was influenced by the past interactions between NGC~4696 and the other galaxies in the Centaurus cluster. 
In Fig.~\ref{fig:E1 profile}-Fig.~\ref{fig:SP2 profile} we show the preliminary results for the radial surface brightness profiled of the other $5$ galaxies that were subtracted from the images.

\newpage
\section{Fixed versus variable color cut}
\label{sec:color cut}

\begin{figure}
    \centering
    \includegraphics[scale=0.18]{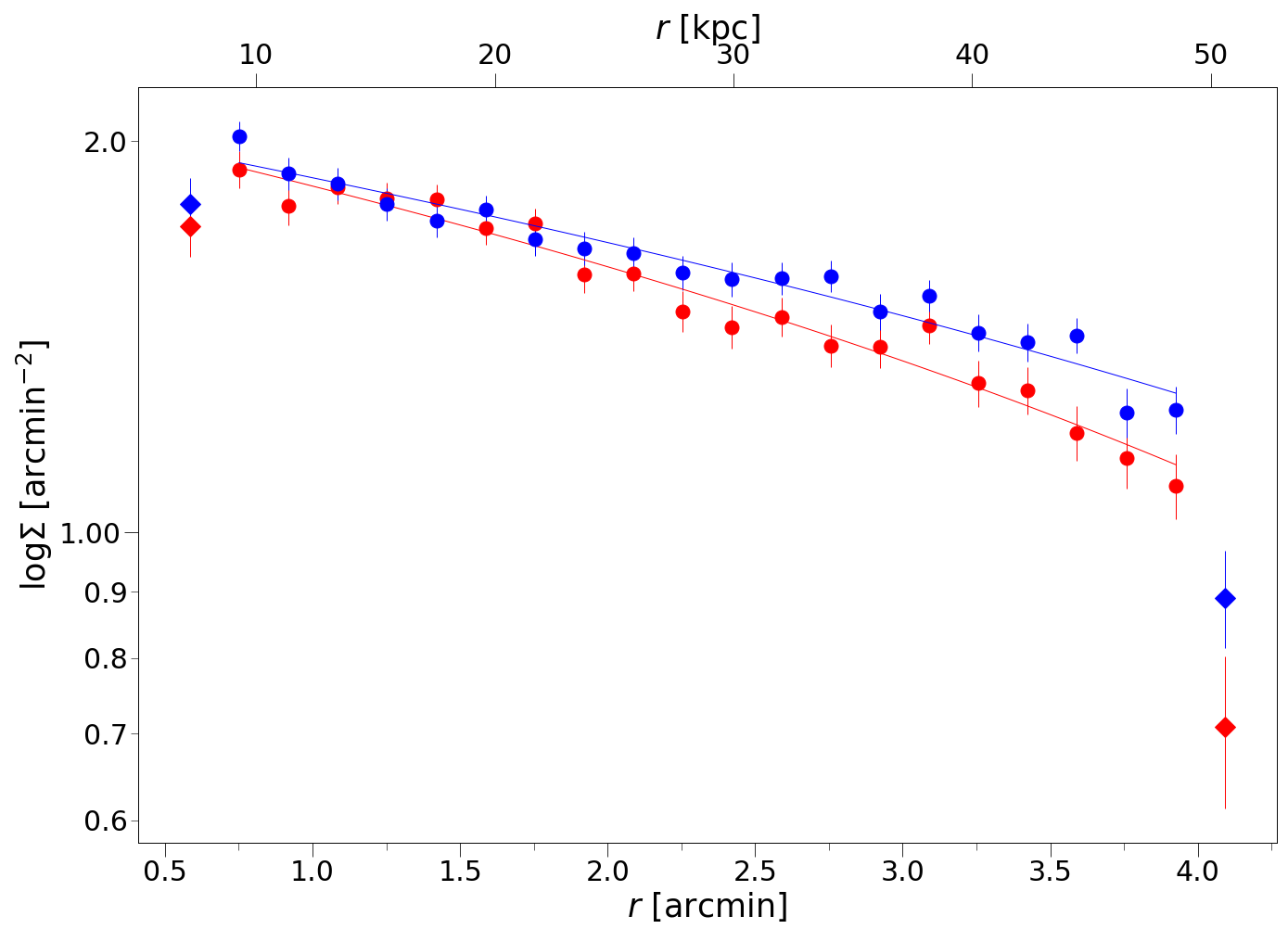}
    \caption{Radial number density profiles for the blue and red GC populations, in the case of a variable $(g'-i')_0$ color separation between the blue and red GC populations. The plot show the logarithm of the density of clusters ($\log\Sigma$) as a function of the galactocentric distance ($r$), for the blue and red populations defined according to Eq.~\ref{eq:blue} and Eq.~\ref{eq:red}. The diamond symbols show the innermost and outermost bins that were not included in the fit, whereas the lines show the best linear fits.}
    \label{fig:fit lim}
\end{figure}

As explained in Sec.~\ref{sec:color distr}, the color distribution shows a trend with radius, with the peaks of both the blue and red populations becoming bluer as we go toward larger galactocentric distances. Moreover, as shown in Tab.~\ref{tab:linear fit} in the two inner radial bins the GC population is better described by a unimodal color distribution, with peaks at $(g'-i')_0=0.921$~mag, whereas in the outer bins the distribution is clearly bimodal. In order to take into account for this radial trend, we reperformed the analysis of the radial number density profile and of the azimuthal distribution using a variable color limit to separate the blue and red GC populations. In particular, we defined as blue GCs those for which:\begin{equation}
    (g'-i')_0<-0.01556*r+0.92990~,
    \label{eq:blue}
\end{equation}
and as red GCs those for which:\begin{equation}
    (g'-i')_0\geq -0.01556*r+0.92990~,
    \label{eq:red}
\end{equation}
where $r$ is the radius, and the numerical values are those found by performing the linear fit on the median of the peaks in the different radial bins (the fit is shown by the green line in Fig.~\ref{fig:linfit_rad}).
With this definition, we obtained a number of $N_{\rm red}=1393$ and $N_{\rm blue}=1690$ for the red and blue GC populations, respectively. We then divided the two samples in radial bins of $r=0.167$~arcmin, as in the previous analysis, and calculated the number density of GCs in each annulus. The results are shown in Fig.~\ref{fig:fit lim} and the best fit parameters are in Tab.~\ref{tab:fit lim}.

Finally, we divided the two samples in bins of $30^{\circ}$ to study the azimuthal distribution, and perform a sinusoidal fit as explained in Sec.~\ref{blue vs red}. The best fit parameters are reported in Tab.~\ref{tab:pa lim}. In order to calculate the parameters of the red and blue GCSs, we used again the output of the 2D Kernel Density Estimation to fit a Bivariate Gaussian Distribution. The positions of the center of the systems, the position angles, minor and major axis, and ellipticities are reported in Tab.~\ref{tab:ell lim}. 
As can be seen, the choice of a variable $(g'-i')_0$ value for the division between the red and blue populations results in a steeper relation for the blue radial number density profile, whereas the red one becomes less steep. Moreover, the peaks of both the red and blue populations are almost identical to what we found using a fixed color separation between the blue and red populations.

\begin{table}[]
\caption{Linear interpolation of the radial density profile of the GC candidates.}
    \centering
    \begin{tabular}{c c c}
    \hline
     & $m$ & $q$  \\
     \hline
    Red & $-0.246\pm 0.011$ & $ 2.093\pm 0.028$ \\ 
    Blue &  $-0.203\pm 0.011$ & $2.078\pm 0.027$ \\ 
    \hline
    \end{tabular}
    \tablefoot{Best-fit parameters for the linear interpolation of the radial density profile of GC candidates in the $\log\Sigma-r$ plane. The lines show the values for the red and blue populations, in the case with a variable $(g'-i')_0$ for the separation between the two populations. $m$ is in arcmin.}
    \label{tab:fit lim}
\end{table}

\begin{table}[]
\caption{Azimuthal distribution in the variable color cut case.}
    \centering
    \resizebox{\columnwidth}{!}{\renewcommand{\arraystretch}{1.5}\begin{tabular}{c c c c c c c c}
    \hline
     & $a$ & $b$ & $c$ & $m$ & $y_{\rm peak}$ & PA$_1$ & PA$_2$  \\
    \hline
    Red$_{\rm Eq.\ref{eq:asym}}$ & $116.09\pm 0.51$ & $19.47^{+0.70}_{-0.74}$ & $52.72^{+0.83}_{-0.81}$ & $2$ & $135.54^{+1.06}_{-1.11}$ & $110.40^{+2.33}_{-1.31}$ & $290.78\pm 0.13$ \\
    Blue$_{\rm Eq.\ref{eq:asym}}$ & $140.81^{+0.76}_{-0.71}$ & $15.18^{+1.07}_{-1.11}$ & $42.59\pm 1.04$ & $2$ & $155.97^{+1.46}_{-1.54}$ & $100.64^{+1.17}_{-2.46}$ & $280.29\pm 0.29$ \\
    Red$_{\rm Eq.\ref{eq:sin}}$ & $116.05^{+0.51}_{-0.50}$ & $19.20^{+0.78}_{-0.75}$ & $29.89^{+0.91}_{-0.93}$ & $-$ & $135.25^{+1.05}_{-1.10}$ & $74.90^{+0.86}_{-1.16}$ & $254.90^{+0.86}_{-1.16}$ \\
    Blue$_{\rm Eq.\ref{eq:sin}}$ & $140.84^{+0.72}_{-0.76}$ & $11.93^{+1.21}_{-1.26}$ & $17.76^{+2.19}_{-2.11}$ & $-$ & $152.77^{+1.54}_{-1.53}$ & $62.76^{+1.89}_{-2.15}$ & $242.76^{+1.89}_{-2.15}$ \\
    \hline
    \end{tabular}}
    \tablefoot{Best-fit parameters and peak values of the sinusoidal interpolation performed with Scipy for the red and blue GC samples, in the case with a variable $(g'-i')_0$ for the separation between the two populations. The errors on the parameters were obtained via the Bootstrap method.}
    \label{tab:pa lim}
\end{table}

\begin{table}
\caption{Shape of the GCS in the variable color cut case.}
    \centering
    \resizebox{\columnwidth}{!}{\renewcommand{\arraystretch}{1.5}\begin{tabular}{c c c c c c c}
    \hline
     & RA & DEC & $a$ & $b$ & PA & $e$  \\
     & (J2000) & (J2000) & (arcmin) & (arcmin) &  & \\
     \hline
    Red & $12^{\rm h}48^{\rm m}49^{\rm s}.61^{+0.30}_{-0.34}$ & $-41^{\circ}18'35''.60^{+5.02}_{-3.74}$ & $3.645^{+0.062}_{-0.058}$ & $2.886^{+0.054}_{-0.059}$ & $90^{\circ}$ & $0.208\pm 0.021$ \\
    Blue & $12^{\rm h}48^{\rm m}49^{\rm s}.10^{+0.31}_{-0.35}$ & $-41^{\circ}18'38''.25^{+3.95}_{-3.55}$ & $3.728^{+0.078}_{-0.063}$ & $2.904^{+0.064}_{-0.052}$ & $90^{\circ}$ & $0.221^{+0.020}_{-0.019}$ \\
    \hline
    \end{tabular}}
    \tablefoot{Parameters of the elliptical shape of the GCS for the red and blue populations, in the case of a variable $(g'-i')_0$ limit between the two populations. The columns report the coordinates of the center of the ellipse (RA and DEC), the major and minor axis ($a$ and $b$), the position angle (PA), and the ellipticity ($e$). The errors were calculated via the Bootstrap method.}
    \label{tab:ell lim}
\end{table}

\section{GC luminosity function and specific frequency}
\label{GCLF}

The globular cluster luminosity function (GCLF) is an important tool to determine the richness of a GC system. Furthermore, its turnover magnitude is used as a distance estimator to its host galaxy (e.g., \citealt{Rej}). It is defined as:\begin{equation}
    \frac{dN}{dM}\propto \frac{1}{\sigma\sqrt{2\pi}}\exp{-\frac{(m-M)^2}{2\sigma^2}}~,
\end{equation}
where $dN$ is the number of GCs in the magnitude bin $dm$, $m$ and $M$ are the apparent and absolute Turnover magnitudes, and $\sigma$ is the width of the Gaussian distribution.

In this section we estimate the distance to NGC~4696 from the GCLF and based on that distance, determine the specific frequency $S_N$ of the GCS. $S_N$ is related to the total number of GCs and to the luminosity of the host galaxy in the $V$ band in units of $M_V=-15$, and it is given by (\citealt{Harris81}):\begin{equation}
    S_N=N_{\rm GC}*10^{0.4*(M_V + 15)}~,
\end{equation}
where $N_{\rm GC}$ is the total number of GCs and $M_V$ is the absolute magnitude of the host galaxy. It has been shown that the specific frequency depends on galaxy type and luminosity, with the highest specific frequencies occurring in low-mass dwarf galaxies and in giant ellipticals in galaxy clusters, where it reaches values $\sim 10$ for $M_V<-20$~mag (\citealt{Peng08}, \citealt{Georgiev10}).

The artificial stars experiment (Sec.~\ref{sec:cmd}) showed that there is a difference between the input magnitudes and those recovered by SExtractor, that increases going toward the fainter magnitudes. In Fig.~\ref{fig:comparison addstar}, the difference difference between the input and output magnitudes in the $i'$ band is shown as a function of the input magnitude corrected for the zeropoint and extinction coefficient. The dots represent the mean of the difference for the artificial stars recovered for a certain input magnitude. This was used to correct the magnitudes for the calculation of the luminosity function.

\begin{figure}
    \centering
    \includegraphics[scale=0.24]{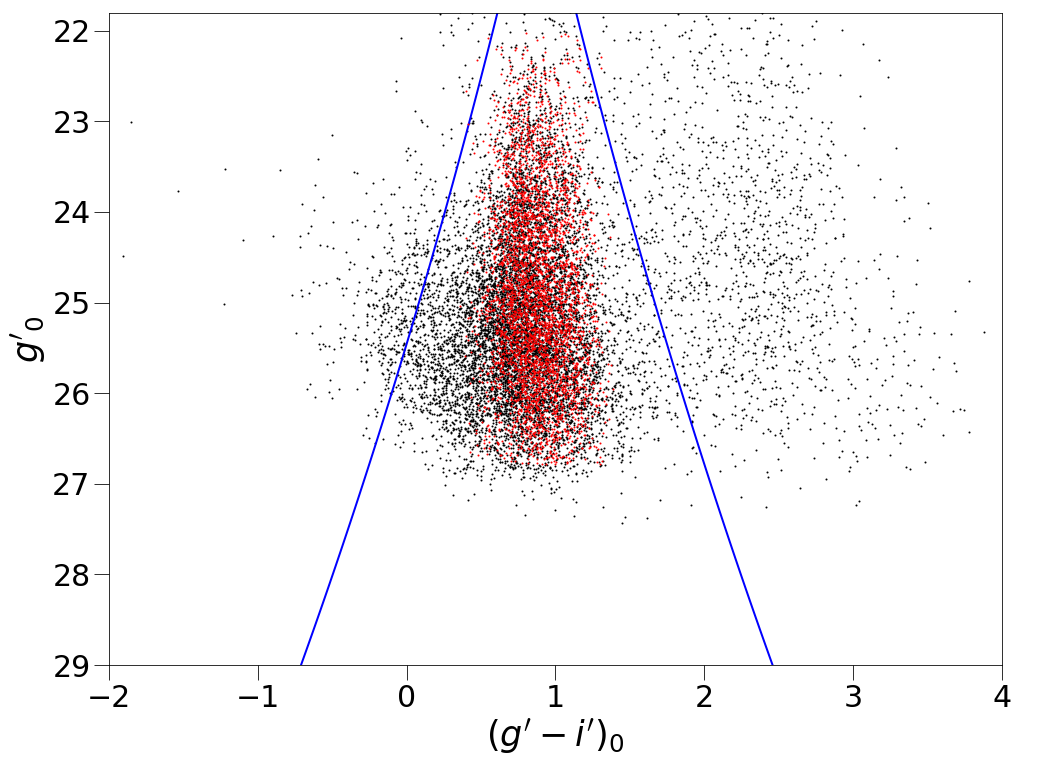}
    \caption{Color-magnitude diagram of the GC candidates, zooming into the plot shown in Fig.~\ref{fig:cmd_class}. The black dots represent the full sample of point-like sources in our field. The red dots are the $3818$ GC candidates defined by the two-color and magnitude selection applied for the analysis in previous sections. The blue lines represent the more inclusive single color $(g'-i')_0$ selection that we adopt for the analysis of the GCLF, see text for details. This corresponding sample yields $6755$ candidates for the calculation of the GCLF.}
    \label{fig:limit exp}
\end{figure}

For the GC selection in the previous sections, we applied two simultaneous color cuts: $0.35<(g'-r')_0<0.85$ and $0.0<(r'-i')_0<0.55$, following \citet{Faifer17}. This selection ensured a clean sample. At the same time, as can be seen from Fig.~\ref{fig:limit exp}, this simultaneous color cut reduces the number of selected GCs at faint magnitudes $g'_0 \gtrsim 25.5$~mag, compared to a sample where only $(g'-i')_0$ color selection would be done. The photometric errors at these fainter magnitudes broaden the color distribution.

For the analysis of the total number of GCs in the GCLF, we therefore adopt a more inclusive GC selection than the two-color selection by \citet{Faifer17}. We apply a single color selection in $(g'-i')_0$ only, and broaden the selection window as a function of magnitude according to the photometric uncertainties as determined from the artificial star experiments. The resulting selection window is indicated in Fig.~\ref{fig:limit exp} and is defined by the following function:
\begin{equation}
    a-\exp{(b*g'_0)}\leq(g'-i')_0\leq c+\exp{(b*g'_0)}~
\end{equation}
where $a=3.619\pm 0.165$, $b=0.051\pm 0.002$, and $c=-1.869\pm 0.165$. Finally, we considered only sources with magnitude $g'_0>22.0$~mag, and in the region closer to NGC~4696 ($r\leq 4.091$~arcmin). This leaves us with a sample of $6755$ GCs. 

To derive the GC luminosity function, we calculated the number of clusters in bins of $0.2$~mag and corrected them according to the completeness values as determined from artificial star experiments (Fig.~\ref{fig:completeness}). We note that those completeness tests were done assuming the average color of the GC sample, and requiring that a source be detected in all three filters. We did not apply an additional two-color selection of the detected sources. This means that those completeness numbers can be applied to the more inclusive sample constructed in this section.

\begin{figure}[h!]
    \centering
    \includegraphics[scale=0.26]{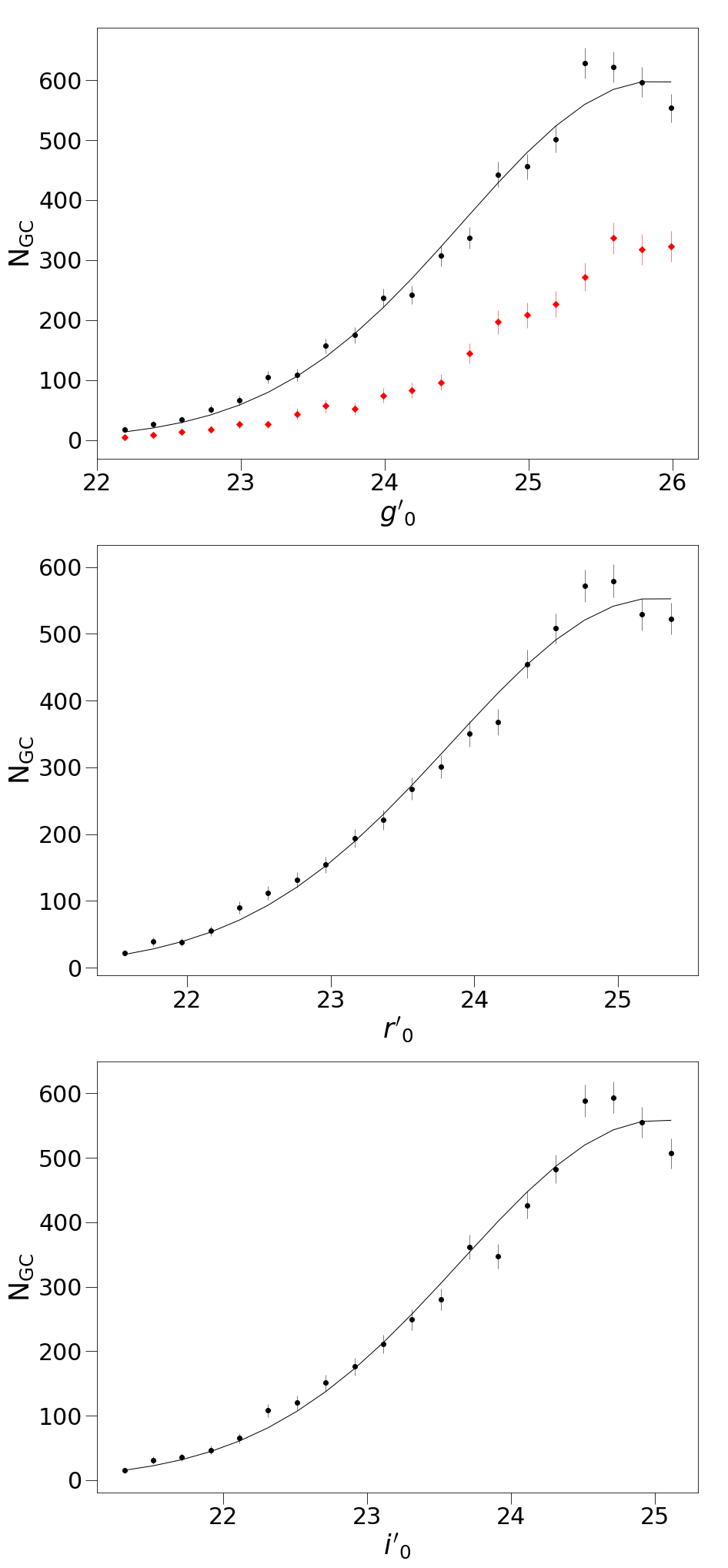}
    \caption{Globular cluster luminosity function. {\em Top panel}: GCLF in the $g'$ band, where the black points represent the number of clusters corrected for the completeness fraction divided in bins of $0.2$~mag, the solid line represent the best fit Gaussian. The red diamonds represent the number of sources in the two rectangular regions considered for the background, corrected by the completeness fraction and normalized by the area. {\em Middle and Bottom panels}: GCLF in the $r'$ and $i'$ band, where the points represent the number of clusters corrected for the completeness fraction divided in bins of $0.2$~mag, and the blue lines correspond to the magnitude of the completeness limit.}
    \label{fig:LF}
\end{figure}

We assumed the absolute turnover magnitude in the in the three bands to be $M_{g'}=-7.2$~mag, $M_{r'}=-7.56$~mag, and $M_{i'}=-7.82$~mag (\citealt{jordan}).
The GCLF assumes the form of a Gaussian distribution, such as: \begin{equation}
    N_{\rm GC}=A\exp{\left[-\frac{1}{2}\left(\frac{x-\mu}{\sigma}\right)^2\right]}~,
\end{equation}
where $A$ is the amplitude, $x$ is the magnitude in the considered band, $\mu$ is the mean that will be used as the turnover magnitude, and $\sigma$ is the standard deviation. 
The result of the Gaussian fit in the $g'$, $r'$ and $i'$ bands are shown in Fig.~\ref{fig:LF}, whereas the magnitudes up to which the fit was performed, the best fit parameters and the calculated distances are reported in Tab.~\ref{tab:GCLF}. The derived distances in the three bands agree with each other within their error bars. The error weighted mean distance between the three bands is $38.36\pm 2.49$~Mpc. This is consistent with the distance of $42.5\pm 3.2$~Mpc from \citet{mieske}. 

As explained in Sec.~\ref{sec:color distr}, we did not apply any statistical correction to our sample for foreground stars since the Besan{\c{c}}on Model of the Galaxy shows that the contribution of foreground stars in this region is negligible. As a further sanity check, we calculated the luminosity function in two rectangular regions with an area of $6$~arcmin$^2$ each,  located in the upper and lower part of the image well outside the region defined by the model of the surface brightness distribution of the galaxy. We applied the same inclusive color selection defined in Fig.~\ref{fig:limit exp} for this comparison field. Concerning background galaxies, we expect the contribution to the background level to be negligible. In fact, as shown in Fig.~3 in \citet{Fensch}, most of the background contributes in the bluest part of the color distribution, which was in part excluded by the considered color selection. Blue background sources could still affect our sample, so in order to check how they would affect the results we made a more strict $(g'-i')_0$ color selection, considering a broadening in color corresponding to $75\%$ of the previous one. This excludes more sources at the blue end of the luminosity function. The number of GCs decreases of $2.26\%$, which results in a nonsignificant change in the values of the distance and specific frequency. 

As a final check, we compared the luminosity function in this comparison field with that for the GC candidates of NGC~4696 (top panel of Fig.\ref{fig:LF}). The counts in the comparison field are renormalized to correspond to the same total area as the NGC~4696 sample. The fit of the luminosity function to the sources in the comparison field shows a fainter turnover magnitude ($g’_0=26.32\pm 0.22$) with respect to the central region of NGC 4696 ($g'_0=25.88\pm 0.14$). In order to quantify the potential effect of faint background sources on top of the GCLF in the comparison field, we forced the fit of the luminosity function with the same $\mu$ and $\sigma$ used for the GC candidates but with free $A$ on the background point-like sources, and excluding the three faintest bins. This showed an excess of background sources in the three lower magnitude bins, corresponding to about $10\%$ of the number counts in the background field. Given the scaling in source density between the background field and the GC candidate field, this implies about $6\%$ contamination by background sources for the GC candidate sample. Based on this, a correction will be applied to the calculation of the specific frequency. In a future work we plan to improve the background level estimation by adding NUV and/or NIR photometry.

The total number of clusters is given by\begin{equation}
    N_{\rm GC}=\frac{\sqrt{2\pi}A*\sigma}{bin ~size}~,
\end{equation} 
where $A$ and $\sigma$ are the amplitude and the standard deviation of the GCLF, respectively, and the $bin ~size=0.2$~mag. The total number of sources is obtained by integrating the fitted GCLF, resulting in $N_{\rm GC,g'}=10069$, $N_{\rm GC,r'}=9948$, $N_{\rm GC,i'}=9741$, respectively. We adopt the mean of those three numbers $N_{\rm GC}=9919$. As apparent magnitude for NGC~4696 we assume the value of $m_V=10.05$, found by \citet{Misgeld} via the curve of growth analysis on VLT images with a field of view smaller than that considered in this work. Assuming a distance modulus of $(m-M)=32.92$ ($38.36$~Mpc as determined previously), we found  $M_{\rm V, NGC~4696}=-22.87$~mag.  With these values, we obtain a lower limit for the specific frequency of $S_N=7.1\pm 0.9$. Correcting this for the estimated $6\%$ background contamination, the final value becomes $S_N=6.8 \pm 0.9$. This agrees within the errorbars with the value of $7.3\pm 1.5$ found in \citet{mieske}, and is also in reasonable agreement with the values obtained for other giant ellipticals (\citealt{Peng08}; \citealt{Georgiev10}; \citealt{Harris2013}), if slightly at the lower end.

%\clearpage

\begin{table*}
\caption{Best-fit parameters for the GCLF.}
    \centering
    \begin{tabular}{c c c c c c c}
    \hline
    Band & mag & $A$ & $\mu$ & $\sigma$ & $M_{\rm TO}$ & $d$  \\
     & (mag) & & (mag) & (mag) & (mag) & (Mpc) \\
    \hline
    $g'$ & $26.589$ & $598.89\pm 17.37$ & $25.88\pm 0.13$ & $1.34\pm 0.09$ & $-7.2\pm 0.20$ & $41.39\pm 4.54$ \\
    $r'$ & $26.166$ & $553.62\pm 14.74$ & $25.27\pm 0.13$ & $1.43\pm 0.08$ & $-7.56\pm 0.20$ & $36.81\pm 4.01$ \\
    $i'$ & $25.911$ & $558.95\pm 19.74$ & $25.04\pm 0.16$ & $1.39\pm 0.11$ & $-7.82\pm 0.20$ & $37.35\pm 4.41$ \\
    \hline
    \end{tabular}
    \tablefoot{The columns show the band, the magnitude up to which the LF was fitted (mag), the amplitude ($A$), the mean ($\mu$), and the standard deviation ($\sigma$). The last two columns show the absolute turnover magnitude for the filter considered and the calculated distance of NGC~4696.}
    \label{tab:GCLF}
\end{table*}

\section{Discussion and conclusions}
\label{sec:conclusion}
We presented the analysis of the GCS of the giant elliptical NGC~4696. The measurements are based on deep \textit{Magellan} 6.5m/MegaCam ($g'$, $r'$, $i'$) photometry. The following results were obtained:\begin{itemize}
    \item Applying a two color selection we identify a total of $3818$ GC candidates. The GC system has a bimodal color distribution, with peaks at $(g'-i')_0=0.763$~mag and $(g'-i')_0=1.012$~mag, with the blue and red populations divided at $(g'-i')_0=0.905$~mag.
    \item It is worth noticing that the blue and red peaks of the distribution appear well separated only in half of the magnitude bins analyzed. In particular, we notice that in the brightest bin the distribution becomes unimodal as in some of the galaxies analyzed by \citet{Faifer11}. It is also interesting to notice that, as shown in Tab.~\ref{tab:GMM}, in the faintest magnitude bins the color distribution changes as well from a bimodal to a unimodal or trimodal shape. In the last considered bin, the number of GC candidates is much lower than in the other ones. As a test in order to have a better number statistic, we joined the two faintest magnitude bins ($g'_0>25.944$~mag). Here we have a total of $616$ sources. We performed the GMM fit and the AIC test. We found that in this subsample a trimodal distribution is preferable to both the unimodal and bimodal distributions. In this case, we found Bootstrap(DD)$=3.53\pm 0.64$ and that the three Gaussians peak at $(g'-i')_0=0.633\pm 0.011$, $(g'-i')_0=0.858\pm 0.005$, and $(g'-i')_0=1.086\pm 0.008$, suggesting then the presence of an intermediate GC population. 
    \item The color distribution of the GC samples, divided according to the galaxy's effective radius, shows a trend with galactocentric radius. In particular, in the outer bins the color distribution present a bimodal shape, whereas the distribution in the inner bin is unimodal, with peaks somewhere in between those of the bimodal cases. This suggests the presence of an intermediate GC population, and could indicate a different origin for these GCs, like accretion of dwarf galaxies and merging events such as was found in previous studies of GCS around early type galaxies (\citealt{caso17}). 
    \item Assuming coeval and old ages of all GCs, we calculated the values of $[\rm{Fe/H}]$ using the conversions described in \citet{Faifer11}, and fitted the corresponding values of $[\rm{Fe/H}]$ for the median peaks with $\log(r/r_{\rm eff})$. We found that the best fit parameters for the linear interpolation are in good agreement with those found by \citet{Harris23}. Moreover, the metallicity distribution appears bimodal with peaks at $[\rm{Fe/H}]=-1.363\pm 0.010$ and $[\rm{Fe/H}]=-0.488\pm 0.012$. 
    \item The radial density profile, limited to a radius of $0.585\leq r\leq 4.091$~arcmin in order to avoid incompleteness effects in the innermost and outermost regions of the galaxy, for the blue and red GC samples (Fig.~\ref{fig:blue red linear fit}) shows different slopes. The plots show that there is a significant difference between the two profiles in the outer regions of the galaxy, with the number of blue clusters being a factor of $2$ higher than the red ones, whereas in the inner parts the numbers are very similar. 
    \item The azimuthal distributions for the total, red and blue populations of GC candidates, limited to a radius $0.585<r<4.091$~arcmin ($0.344~r_{\rm eff}<r<2.406~r_{\rm eff}$), are well fitted by an asymmetrical sinusoidal distribution. In particular, we found that the peaks of the distributions are close or almost perfectly aligned to the angular positions of the galaxies NGC~4709 and NGC~4696B. Moreover, we found that the ellipticity of the GCS is different with respect to that of the host galaxy. This indicates that the interactions between NGC~4696 and these two galaxies played a role in shaping the GCS under consideration. Moreover, these results agree with the merger/interaction history outlined by previous works on the hot X-ray gas, such as the identification of a filamentary structure connecting NGC~4696 with NGC~4696B and a metallicity excess around NGC~4709 by \citet{Walker}, and an asymmetric temperature variation with the hottest region coinciding with NGC~4709 by \citet{Churazov}.
    \item Since we notice that the color distribution shows a trend with radius, we decided to recalculate the radial density profiles and the azimuthal distributions using a variable $(g'-i')_0$ color separation between the blue and red GC populations. In particular, we separated the two using the linear fit on the median between the blue and red peaks (given by $(g'-i')_0=-0.01556*r+0.92990$). In this case, the radial density profile of the blue population becomes steeper than in the case of a constant color separation, and the differences between the two profiles become negligible in the inner regions and much less pronounced in the outer regions. In a future work, we aim at expanding our sample to further investigate the importance of the choice of the color separation between the blue and red GC populations.
    \item We calculated the GCLF in the three filters, and used it for our own estimation of the distance of NGC~4696, as well as the specific frequency. As can be seen from Tab.~\ref{tab:GCLF}, the distances estimated in the different filters are consistent within the errors. Moreover, our results are consistent with the measurement of the distance of the Centaurus cluster obtained with both the GCLF and the surface brightness fluctuation method by \citet{mieske}. Finally, we used the peak of the GCLF to estimate the total number of GCs in NGC~4696, assuming that the luminosity function is described by a symmetric Gaussian. We then used the number of clusters to estimate the specific frequency of the system, and corrected it for the $6\%$ contamination by background galaxies estimated from the comparison between the luminosity function of the GC candidates and that of the sources in the two rectangular regions defined in Sec.~\ref{sec:GCsel}. The specific frequency resulted of $S_N\sim 6.8$. This value is to consider a lower limit since the study is limited to a radius of $4.091$~arcmin, so we are missing the GC population at the highest galactocentric distances. However, it is well within the range for giant ellipticals in galaxy clusters, and agrees with previous calculations (\citealt{mieske}).
\end{itemize} 
As a final note, the next step in the analysis will be the spectroscopic follow-up of our GC candidates. The addition of spectroscopic data to our study will be fundamental to confirm their affiliation to NGC~4696. This will allow also to investigate with more precision the interaction history of the host galaxy, and to understand the history of the cluster itself, as was highlight in other analysis such as that of the Virgo (\citealt{Durrell}), Coma (\citealt{Madrid}) and the Fornax (\citealt{Chaturvedi}) clusters, and Abell~1689 (\citealt{Alamo}). Moreover, we will add the other field to the analysis, starting with that containing NGC~4709. This will allow us to investigate its GCS, as well as the GCs between this galaxy and NGC~4696, in order to determine the role of the interactions in shaping the GCSs of these two galaxies.

\section*{Acknowledgements}

S.~F. and M.~G. gratefully acknowledge support by Fondecyt Project N$^{\circ}~1220724$. \\
This research has made use of the NASA/IPAC Extragalactic Database (NED), which is funded by the National Aeronautics and Space Administration and operated by the California Institute of Technology.

%%%%%%%%%%%%%%%%%%%%%%%%%%%%%%%%%%%%%%%%%%%%%%%%%%
\section*{Data Availability}

All tables are available electronically from the author upon request.
The MegaCam images are accessible through the CADC image archive.



\clearpage
%\begin{thebibliography}{99}
%\end{thebibliography}

%%%%%%%%%%%%%%%%%%%%%%%%%%%%%%%%%%%%%%%%%%%%%%%%%%

%%%%%%%%%%%%%%%%% APPENDICES %%%%%%%%%%%%%%%%%%%%%

\onecolumn
\begin{appendix}

\section{Catalog of the globular cluster candidates}

\begin{table*}[h!]
\caption{Properties of the globular cluster candidates of the galaxy NGC~4696.}
    \centering
    \begin{adjustbox}{width=\textwidth}
    \begin{tabular}{c c c c c c c c}
    \hline
    RA & DEC & CLASS$_{g'}$ & CLASS$_{r'}$ & CLASS$_{i'}$ & $g'_0$ & $r'_0$ & $i'_0$  \\
    (hh:mm:ss) & ($^{\circ}~'~"$) & & & & (mag) & (mag) & (mag) \\
    \hline
  12:48:59.27 & -41:23:35.50 & 0.952 & 0.800 & 0.968 & 25.094$\pm$0.164 & 24.550$\pm$0.132 & 24.126$\pm$0.148\\
  12:48:41.57 & -41:23:31.00 & 0.847 & 0.714 & 0.981 & 24.841$\pm$0.165 & 24.040$\pm$0.134 & 23.564$\pm$0.112\\
  12:48:39.73 & -41:23:30.71 & 0.680 & 0.963 & 0.974 & 24.562$\pm$0.140 & 23.973$\pm$0.113 & 23.645$\pm$0.109\\
  12:48:43.43 & -41:23:23.52 & 0.984 & 0.969 & 0.504 & 24.272$\pm$0.114 & 23.458$\pm$0.092 & 22.908$\pm$0.083\\
  12:48:57.14 & -41:23:20.76 & 0.847 & 0.232 & 0.981 & 24.921$\pm$0.116 & 24.522$\pm$0.093 & 24.393$\pm$0.152\\
  12:49:03.28 & -41:23:19.85 & 0.939 & 0.099 & 0.957 & 25.311$\pm$0.150 & 24.807$\pm$0.120 & 24.685$\pm$0.136\\
  12:48:57.76 & -41:23:17.62 & 0.054 & 0.030 & 0.981 & 23.337$\pm$0.040 & 22.783$\pm$0.032 & 22.753$\pm$0.044\\
  12:48:51.34 & -41:23:17.48 & 0.839 & 0.933 & 0.982 & 23.800$\pm$0.056 & 23.069$\pm$0.045 & 23.024$\pm$0.048\\
  12:48:42.39 & -41:23:19.84 & 0.098 & 0.242 & 0.983 & 23.344$\pm$0.043 & 22.723$\pm$0.034 & 22.711$\pm$0.048\\
  12:48:45.36 & -41:23:19.19 & 0.884 & 0.632 & 0.949 & 25.174$\pm$0.157 & 24.804$\pm$0.125 & 24.763$\pm$0.190\\
  12:48:37.02 & -41:23:15.34 & 0.077 & 0.948 & 0.983 & 23.078$\pm$0.035 & 22.394$\pm$0.029 & 22.215$\pm$0.034\\
  12:48:49.90 & -41:23:11.95 & 0.930 & 0.646 & 0.956 & 25.374$\pm$0.156 & 24.901$\pm$0.125 & 24.366$\pm$0.161\\
  12:48:58.18 & -41:23:14.32 & 0.130 & 0.024 & 0.964 & 24.185$\pm$0.077 & 23.644$\pm$0.062 & 23.197$\pm$0.063\\
  \hline
    \end{tabular}
    \end{adjustbox}
    \tablefoot{The columns report the coordinates (RA and DEC), the stellarity index in each filter (CLASS$_{g'}$, CLASS$_{r'}$, and CLASS$_{i'}$), the extinction-corrected magnitude and their errors ($g'_0$, $r'_0$, and $i'_0$). The complete catalog is available at the CDS.}
    \label{tab:catalog}
\end{table*}

\section{Sources from previous catalogs}

\begin{table*}[h!]
\caption{Properties of the candidates in our sample for which \citet{Mieske07} confirmed the membership to the Centaurus cluster.}
    \centering
    \begin{adjustbox}{width=\textwidth}
    \begin{tabular}{c c c c c c c c c c c c c}
    \hline
    CCOS & RA & DEC & V$_0$ & (V-R)$_0$ & $v_{\rm r}$ & CLASS$_{i'}$ & $g'_0$ & $r'_0$ & $i'_0$ & $(g'-i')_0$ & $(r'-i')_0$ & $(g'-r')_0$ \\
     & (hh:mm:ss) & ($^{\circ}~'~"$) & (mag) & (mag) & (km~s$^{-1}$) & & (mag) & (mag) & (mag) & (mag) & (mag) & (mag) \\
     \hline
    J1248.79-4117.15 & 12:48:47.6 & -41:17:09.9 & 21.1 & 0.5 & 2882$\pm$54 & 0.983 & 21.521$\pm$0.007 & 20.873$\pm$0.004 & 20.508$\pm$0.003 & 1.013 & 0.364 & 0.648\\
  J1248.74-4118.58 & 12:48:44.7 & -41:18:35.3 & 21.76 & 0.44 & 2189$\pm$114 & 0.609 & 22.037$\pm$0.010 & 21.570$\pm$0.006 & 21.333$\pm$0.007 & 0.704 & 0.238 & 0.466\\
  J1248.76-4118.70 & 12:48:45.5 & -41:18:42.4 & 21.77 & 0.55 & 2828$\pm$45 & 0.982 & 22.087$\pm$0.010 & 21.432$\pm$0.005 & 21.098$\pm$0.005 & 0.989 & 0.334 & 0.655\\
  J1248.97-4120.01 & 12:48:58.2 & -41:19:59.7 & 22.12 & 0.52 & 2962$\pm$98 & 0.979 & 22.111$\pm$0.011 & 21.461$\pm$0.006 & 21.141$\pm$0.006 & 0.970 & 0.320 & 0.650\\
  J1248.74-4118.95 & 12:48:44.1 & -41:18:57.1 & 22.17 & 0.53 & 3213$\pm$53 & 0.983 & 22.390$\pm$0.013 & 21.796$\pm$0.007 & 21.550$\pm$0.008 & 0.840 & 0.246 & 0.594\\
  J1248.70-4118.23 & 12:48:42.0 & -41:18:14.4 & 22.18 & 0.48 & 2822$\pm$55 & 0.979 & 22.177$\pm$0.011 & 21.455$\pm$0.006 & 21.100$\pm$0.005 & 1.077 & 0.354 & 0.723\\
  J1248.91-4117.70 & 12:48:54.6 & -41:17:42.4 & 22.29 & 0.43 & 2909$\pm$120 & 0.979 & 22.536$\pm$0.015 & 22.000$\pm$0.008 & 21.727$\pm$0.009 & 0.809 & 0.273 & 0.536\\
  J1248.99-4118.08 & 12:48:59.6 & -41:18:04.7 & 22.31 & 0.53 & 2806$\pm$75 & 0.984 & 22.528$\pm$0.015 & 21.943$\pm$0.008 & 21.680$\pm$0.009 & 0.848 & 0.263 & 0.585\\
  \hline
    \end{tabular}
    \end{adjustbox}
    \tablefoot{The columns report the name of the source (CCOS), the coordinates (RA and DEC), the extinction-corrected V magnitude (V$_0$), the extinction-corrected color index (V-R)$_0$, and the radial velocity ($v_{\rm r}$) obtained by {\citet{Mieske07}}. The last columns show the stellarity index (CLASS$_{i'}$), extinction-corrected magnitudes ($g'_0$, $r'_0$, and $i'_0$), and colors ($(g'-i')_0$, $(r'-i')_0$, and $(g'-r')_0$) obtained in this work.}
    \label{tab:mieske comparison}
\end{table*}

\clearpage

\section{Exposure map}

\begin{figure*}[h!]
    \centering
    \includegraphics[width=\textwidth]{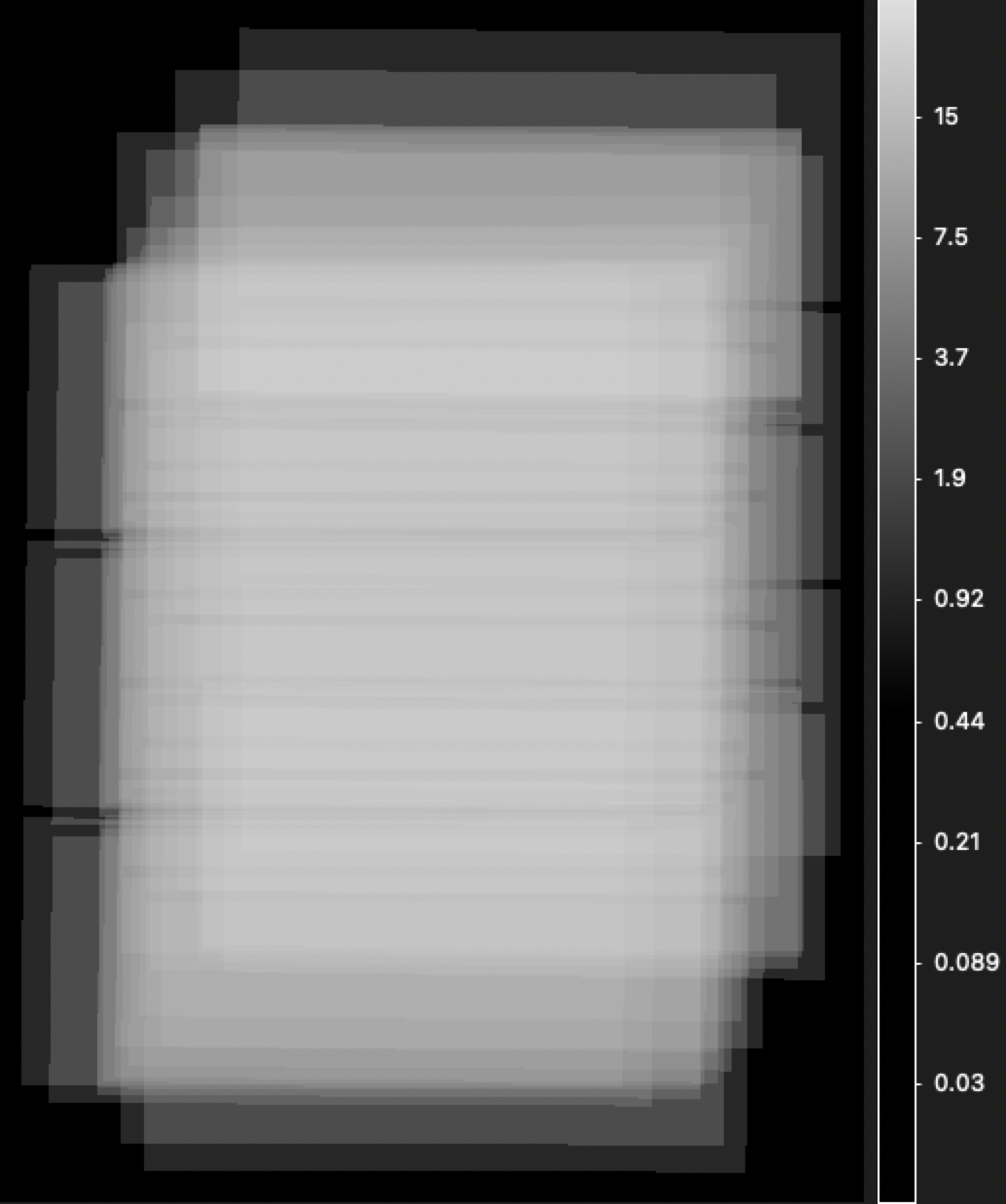}
    \caption{Exposure map in the $r'$ band of the region of the Centaurus cluster analyzed in this work. The colorbar shows the number of participating individual exposures in log scale.}
    \label{fig:exp}
\end{figure*}

\clearpage

\section{The galaxies radial surface brightness profiles}

\begin{figure*}[h!]
    \centering
    \includegraphics[scale=0.35]{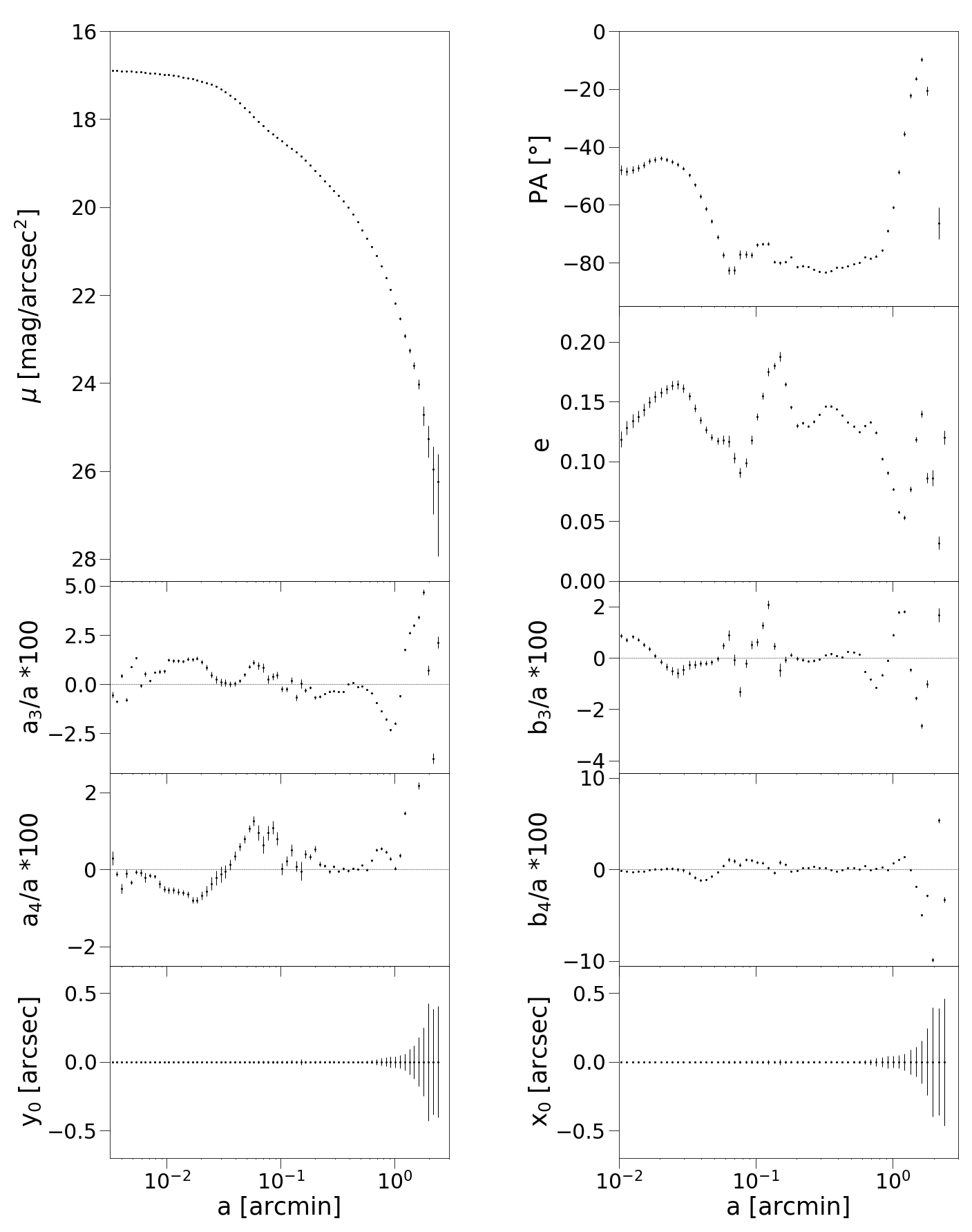}
    \caption{Radial surface brightness profile along the major axis for NGC~4696. The panels show the trend of the surface brightness ($\mu$), the position angle (PA), the ellipticity ($e$), the center ($x_0$,$y_0$), and the Fourier coefficients $a_3$, $a_4$, $b_3$, and $b_4$ as a function of the logarithm of the semi major axis ($a$). The plots were obtained through the interpolation of {\tt ISOFIT} on the image in the $i'$ band.}
    \label{fig:ngc profile}
\end{figure*}

\begin{figure*}
    \centering
    \includegraphics[width=\textwidth]{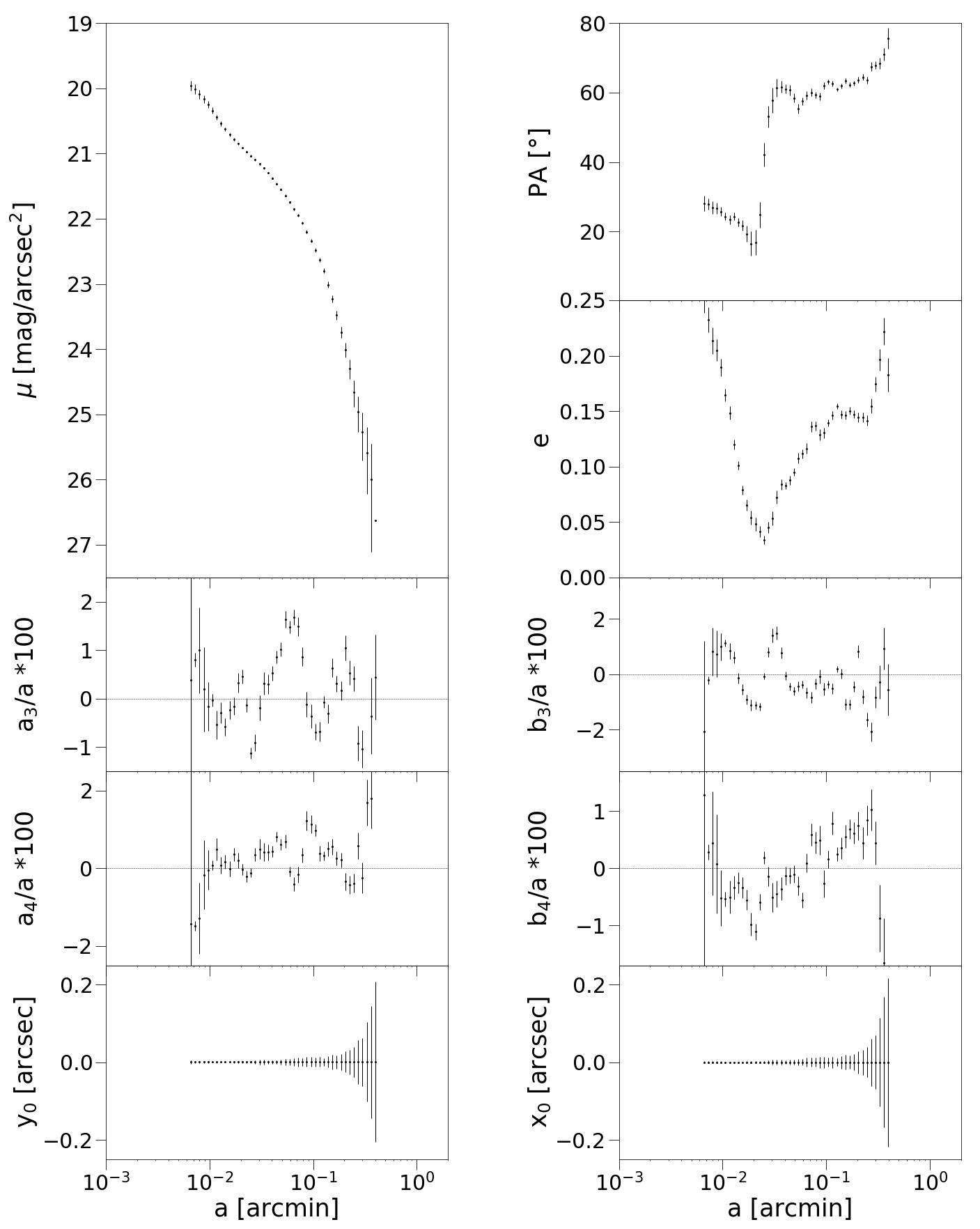}
    \caption{Same as in Fig.\ref{fig:ngc profile}, but for the galaxy WISEA J124839.70-411605.1.}
    \label{fig:E1 profile}
\end{figure*}

\begin{figure*}
    \centering
    \includegraphics[width=\textwidth]{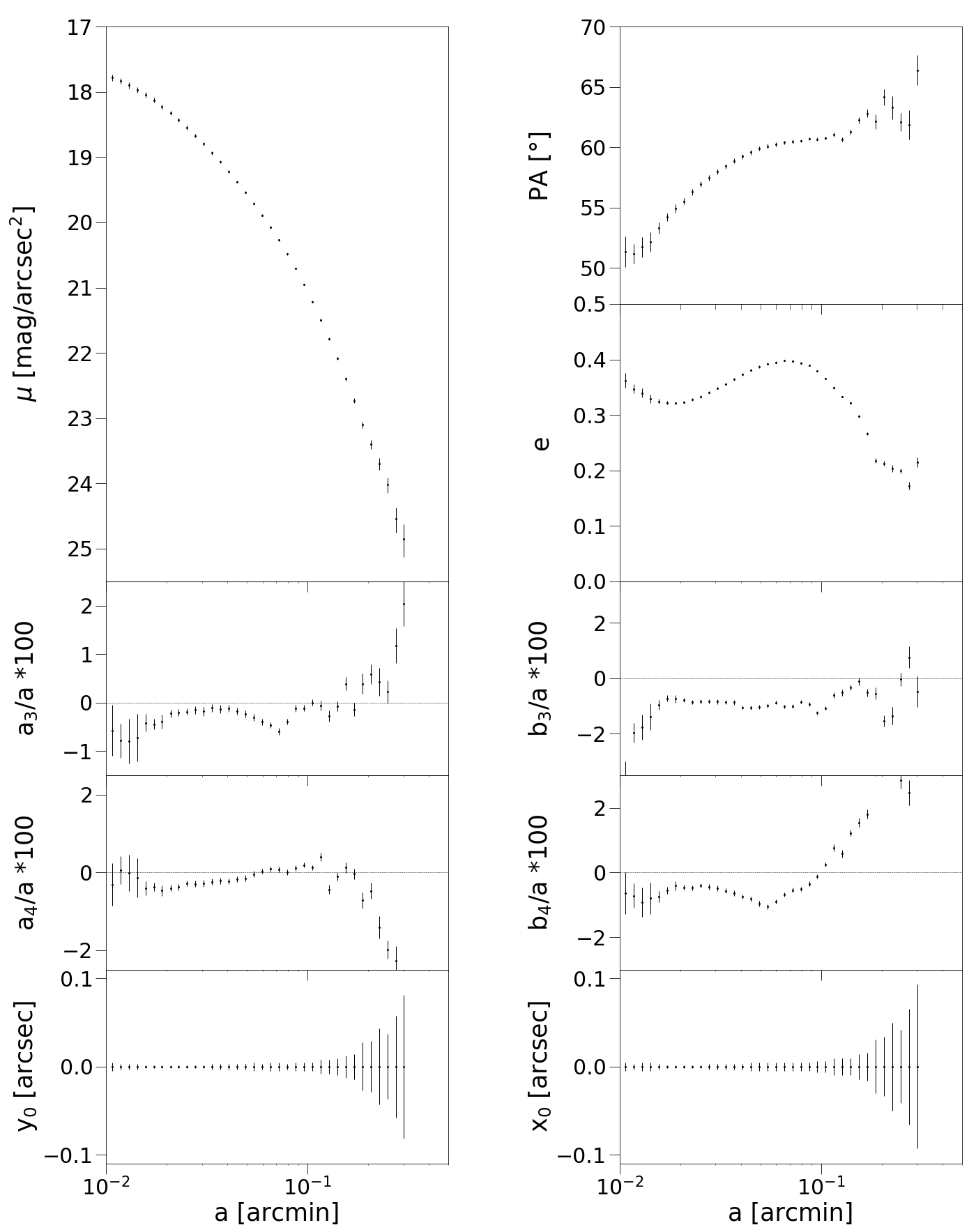}
    \caption{Same as in Fig.\ref{fig:ngc profile}, but for the galaxy WISEA J124831.02-411823.2.}
    \label{fig:E2 profile}
\end{figure*}

\begin{figure*}
    \centering
    \includegraphics[width=\textwidth]{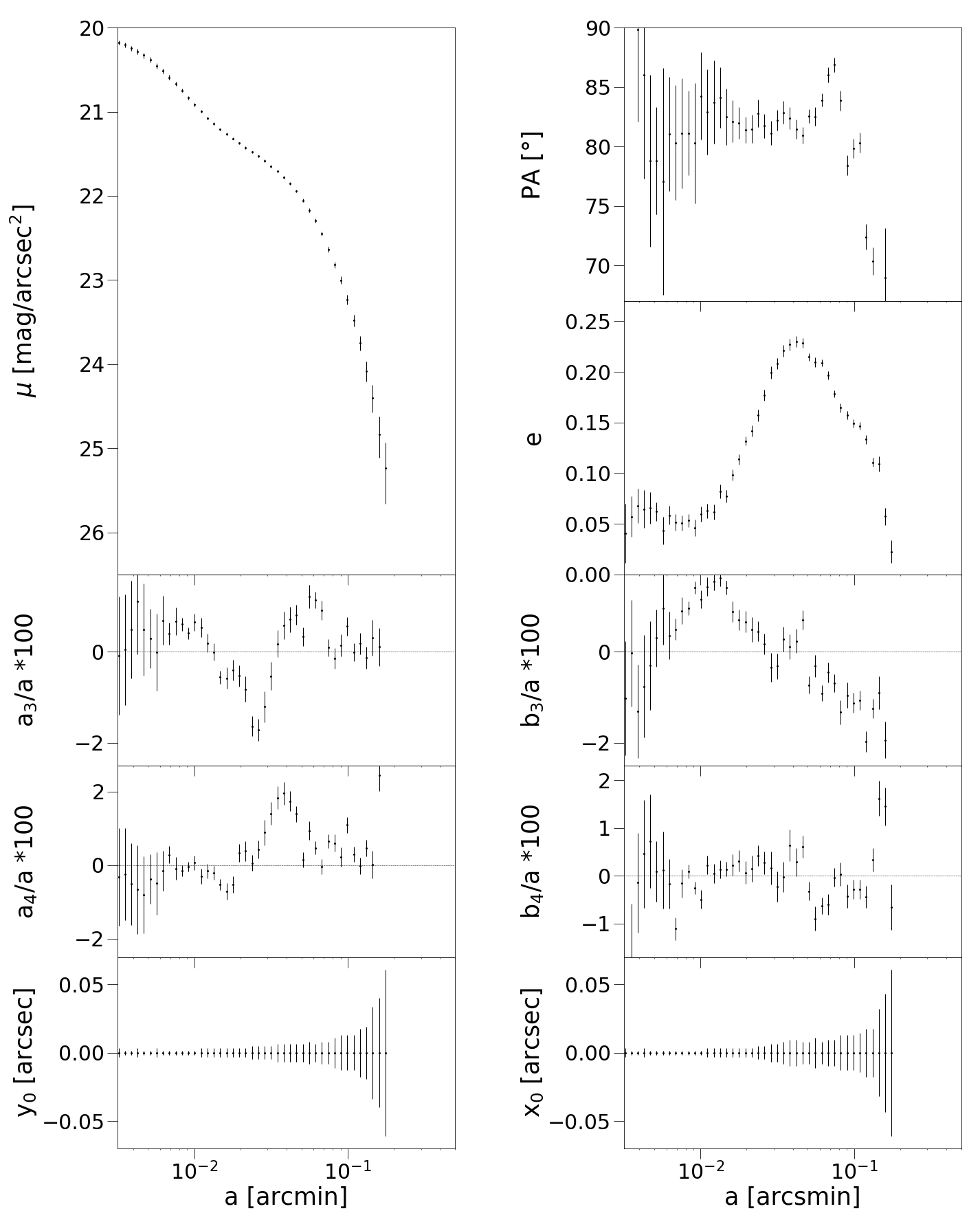}
    \caption{Same as in Fig.\ref{fig:ngc profile}, but for the galaxy WISEA J124902.02-411533.7.}
    \label{fig:E3 profile}
\end{figure*}

\begin{figure*}
    \centering
    \includegraphics[width=\textwidth]{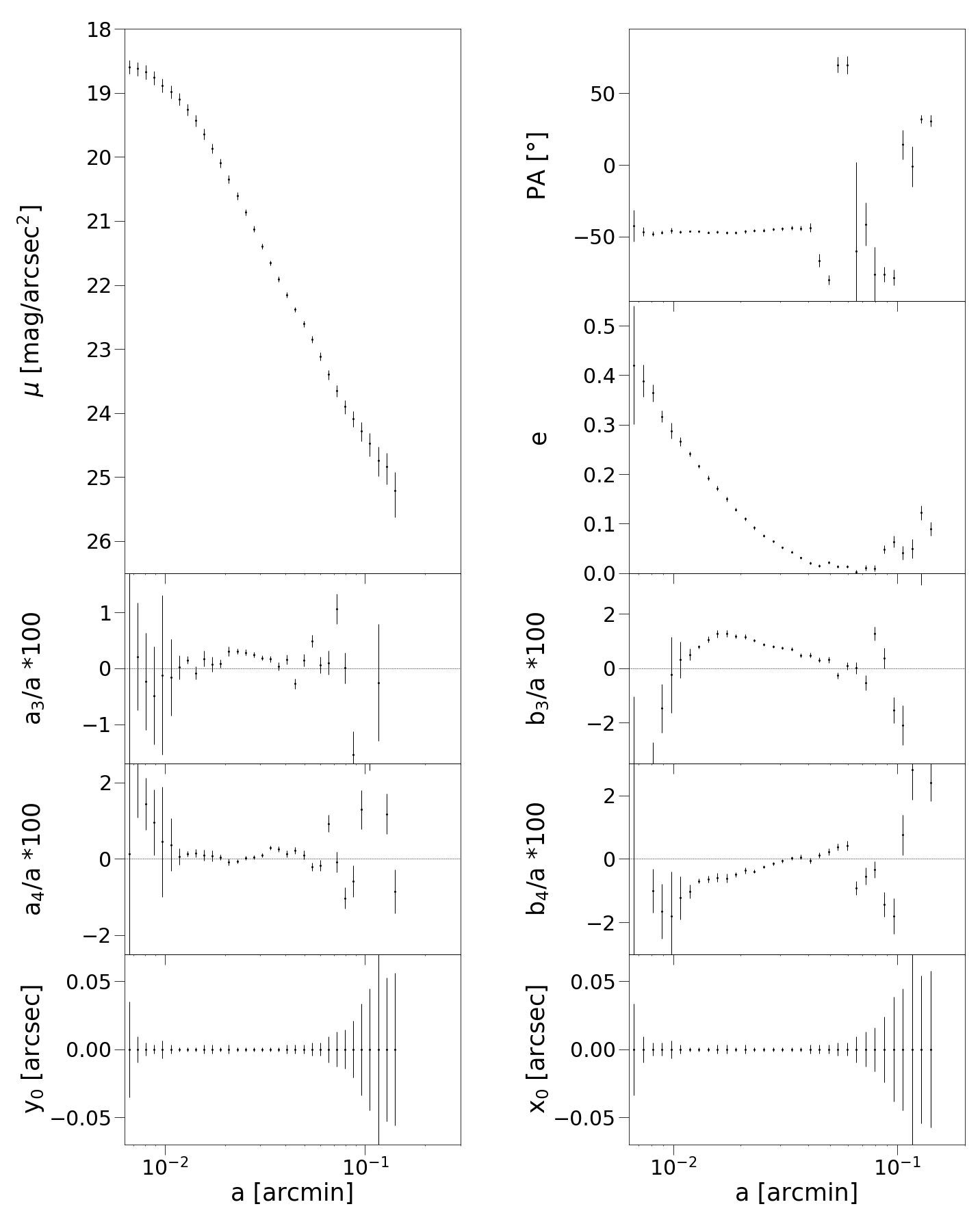}
    \caption{Same as in Fig.\ref{fig:ngc profile}, but for the galaxy WISEA J124846.61-411539.4.}
    \label{fig:SP1 profile}
\end{figure*}

\begin{figure*}
    \centering
    \includegraphics[width=\textwidth]{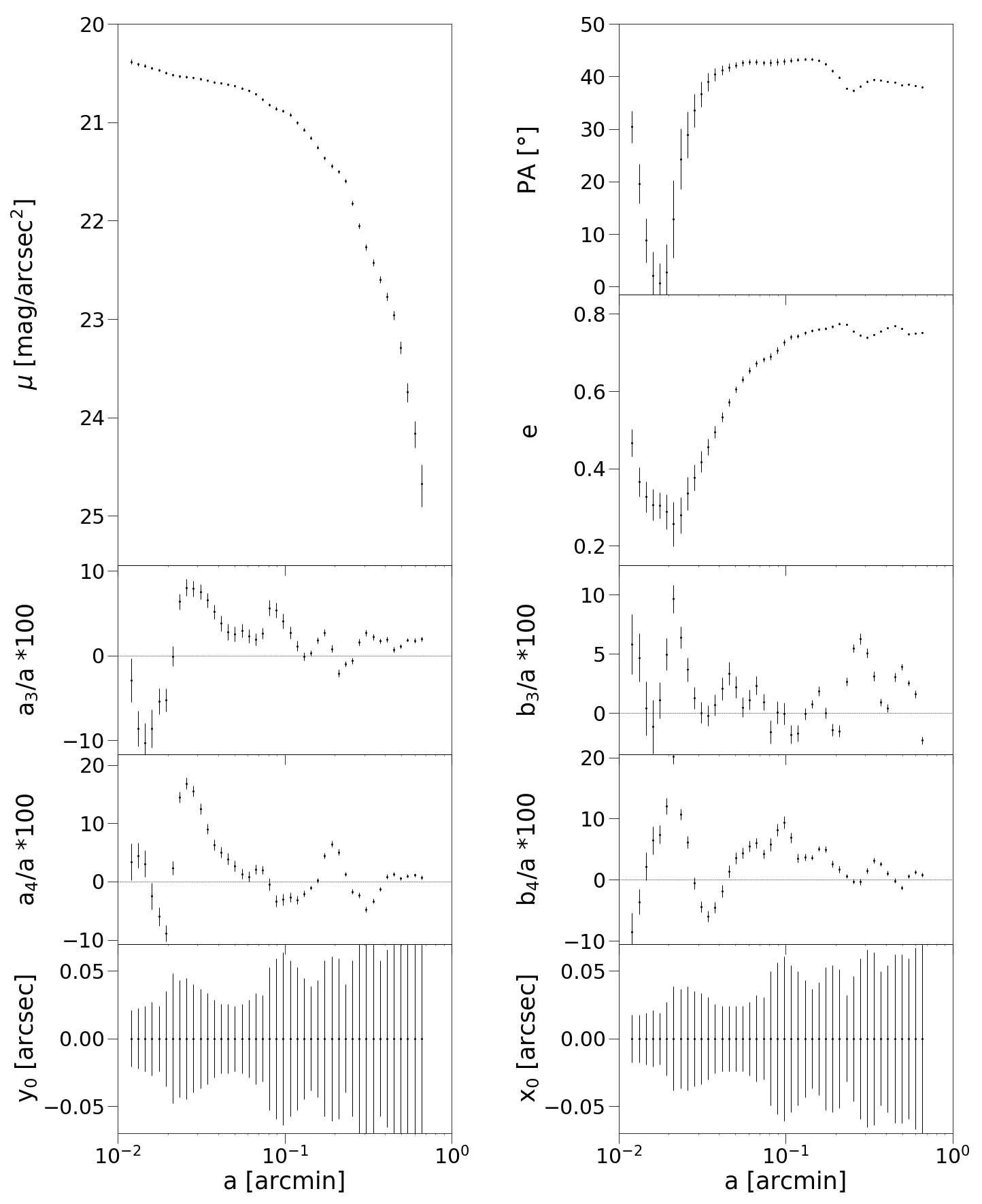}
    \caption{Same as in Fig.\ref{fig:ngc profile}, but for the galaxy ESO 322- G 093.}
    \label{fig:SP2 profile}
\end{figure*}

\clearpage

\section{Comparison between input and output magnitudes for the artificial stars experiment}

\begin{figure}[h]
    \centering
    \includegraphics[width=\textwidth]{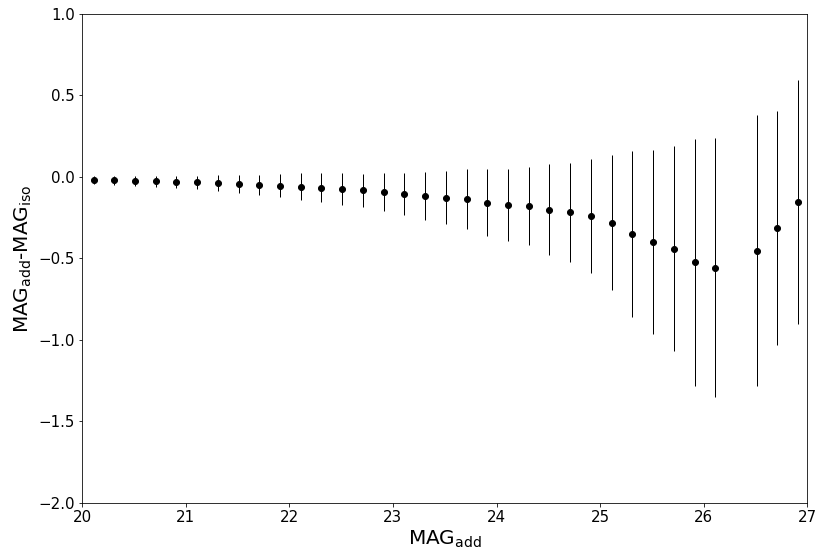}
    \caption{Comparison between the input and output magnitudes for the artificial stars' experiment in the $i'$ band. The dots the means of the difference between the input and output magnitudes ($MAG_{\rm add}-MAG_{\rm iso}$), whereas the errorbars are their RMS.}
    \label{fig:comparison addstar}
\end{figure}

\end{appendix}
%%%%%%%%%%%%%%%%%%%%%%%%%%%%%%%%%%%%%%%%%%%%%%%%%%

\end{document}